\documentclass[11pt,draftcls,onecolumn]{IEEEtran}
\newif\ifpdf
\ifx\pdfoutput\undefined
  \pdffalse
\else
  \ifnum\pdfoutput=1
    \pdftrue
  \else
    \pdffalse
  \fi
\fi
\usepackage{amsmath,epsfig, cite, cases}
\usepackage{amssymb,amscd,mathrsfs}
\usepackage[active]{srcltx}
\usepackage{color}
\usepackage{setspace}
\usepackage{enumerate}
\usepackage{bbding}

\def\E{\mathbb{E}}
\def\alphaN{{\alpha_{M}}}
\def\sig{{\sigma}}
\def\F{{\mathbf{F}}}
\def\Re{{\mathbb{R}}}
\def\xklpf{{\hat{x}^{klpf,M}}}

\def\D{{\mathcal{S}}}

\def\var{n}

\def\GCSNx{{GCSN_{d,n}(x; \mu, \Sigma, D, \D^n, \Delta)}}

\def\Q{{g}}
\def\Qspace{{\mathbb{G}}}

\newcommand{\bra}[1]{\left(#1\right)}

\newtheorem{thm}{Theorem}[section]
\newtheorem{cor}[thm]{Corollary}
\newtheorem{lem}[thm]{Lemma}

\newtheorem{define}{Definition}

\title{\LARGE \bf
The Kalman Like Particle Filter : Optimal Estimation With Quantized Innovations/Measurements*\footnote{* An early version of this paper, with preliminary results, first appeared in \cite{KLPF}}}

\author{Ravi Teja Sukhavasi and Babak Hassibi
\thanks{Ravi Teja Sukhavasi is a graduate student with the department of Electrical Engineering, California Institute of Technology, Pasadena, USA
        {\tt\small teja@caltech.edu}}%
\thanks{Babak Hassibi is a faculty with the department of Electrical Engineering, California Institute of Technology,
        Pasadena, USA
        {\tt\small hassibi@caltech.edu}}%
\thanks{This work was supported in part by the National Science Foundation under grants CCF-0729203, CNS-0932428 and CCF-1018927, by the Office of Naval Research under the MURI grant N00014-08-1-0747, and by Caltech's Lee Center for Advanced Networking.}
}

\begin{document}

\maketitle
\thispagestyle{empty}
\pagestyle{empty}

\begin{abstract}

We study the problem of optimal estimation and control of linear systems using quantized measurements, with a focus on applications over sensor networks. We show that the state conditioned on a causal quantization of the measurements can be expressed as the sum of a Gaussian random vector and a certain truncated Gaussian vector. This structure bears close resemblance to the full information Kalman filter and so allows us to effectively combine the Kalman structure with a particle filter to recursively compute the state estimate. We call the resulting filter the Kalman like particle filter (KLPF) and observe that it delivers close to optimal performance using far fewer particles than that of a particle filter directly applied to the original problem. We show that the conditional state density follows a, so called, generalized closed skew-normal (GCSN) distribution. We further show that for such systems the classical separation property between control and estimation holds and that the certainty equivalent control law is LQG optimal.
\end{abstract}

\section{Introduction}
\label{sec: introduction}

%

Recent advances in very large-scale integration and microelectromechanical system technology have led to the availability of cheap, low quality and low power consumption sensors in the market. This has generated a great deal of interest in wireless sensor networks (WSNs) due to their potential applications in several diverse fields \cite{akyil}. Sensor network constraints such as limited bandwidth and power have inspired a considerable amount of research in developing energy efficient algorithms for network coverage and decentralized detection and estimation using quantized sensor observations \cite{energy, luo, rickard}. Sensor networks also give rise to non-classical information patterns, for example, at any given time, there is no observer in the system that has access to the collective measurements made by all the sensors. One of the many consequences is that, even in the presence of a linear state space structure, the classical Kalman filter would not be applicable.  

The problem of estimation with quantized measurements is almost as old as the Kalman filter itself. An early survey on the subject can be found in \cite{curry_book}. However, most of the earlier techniques centered around using numerical integration methods to approximate the optimal state estimate. The advent of particle filtering \cite{gordon, survey, tutorial} created a whole new set of tools to handle non-linear estimation problems. For example, \cite{rickard} proposes a particle filtering solution for optimal filtering using quantized sensor measurements. But, quantizing sensor measurements can lead to large quantization noises when the observed values are large which then leads to poor estimation accuracy. A natural alternative is to quantize the prediction error.
In \cite{brockett}, this coding technique is referred to as the \lq generalized mean coder-estimator' technique and under a very restrictive state space model, this estimator is shown to be open loop mean-squared stable if the quantizer rate is sufficiently high. The same scheme is independently proposed in \cite{SOI_KF, NTU}, where it is referred to as the \lq sign of innovation' method. Under a symplifying assumption that the prior conditional state density is approximately Gaussian, the optimal filter takes a simple analytical form\footnote{which we refer to as the multiple level quantized Kalman filter (MLQ-KF)} whose error covariance satisfies a modified Riccati recursion (MLQ-Riccati) of the type that appears in a different context in \cite{bruno}. When the state is available at the sensor, \cite{yuksel1} studies an adaptive quantization technique and proves that it can track an unstable process in open loop with a finite mean squared error. 

For linear time invariant dynamical systems, if the Gaussian assumption of \cite{SOI_KF, NTU} were realistic, convergence of the MLQ-Riccati must mean the convergence of the error of the MLQ-KF. \cite{teja} provides examples for which the actual error performance of MLQ-KF does not converge to the MLQ-Riccati which means that the assumption of Gaussianity is not generally true. Therefore, we present a closer examination of the conditional state density in this paper. We derive a novel stochastic characterization of the conditional state density (see Theorem \ref{thm: quant_sto}). A careful literature review reveals that related observations have been made in \cite{curry} and \cite{doucet}. In particular, with some effort, \cite{curry} can be used to derive Theorem \ref{thm: quant_sto} while \cite{doucet} constitutes a special case of the results presented here. Using Theorem \ref{thm: quant_sto}, it is straighforward to see that the conditional state density is not Gaussian. This is to be expected given the non-linear nature of quantization. In fact, it is what we refer to as the Generalized Closed Skew Normal (GCSN) distribution, which is very similar to those studied in \cite{azzalini_multivariate_1996, valle_fundamental_2005, genton_moments, genton_discussion_2005, skewed_kalman, genton_book}. Specializing this result to state space models, we develop a novel particle filtering approach to estimate the state using quantized measurements/innovations and study its asymptotic behavior. Finally, we show that under the information pattern studied, the classical separation property between estimation and control holds for the finite horizon LQG problem. The separation principle has been observed in several settings (for e.g., see \cite{tatikonda2, yuksel5}). It should be noted that for such separation results to be useful in practice, one needs a way to compute the MMSE estimate of the hidden state and this is primarily what we address through this work. 
The proposed filter requires far fewer particles than that of a particle filter applied directly to the original problem \cite{teja}, as will be shown through various simulations. A preliminary version of this work appeared in \cite{KLPF}.

\subsection{Notation}
The following notation will be used in the rest of the paper.
\begin{enumerate}
 \item If $\{u_n\}_{n=-\infty}^{\infty}$ is a discrete time random process, $u_{i:j}$ denotes $\{u_i,\ldots,u_j\}$. 
 \item For random vectors $X$, $Y$, $\langle X, Y\rangle \triangleq E(X-EX)(Y-EY)^T$ and $\Vert X\Vert^2 = \langle X,X\rangle$.
 \item For random variables $(X_1,\ldots, X_n)$, $\mathcal{L}(X_1, \ldots, X_n)$ denotes their linear span. 
 \item $N_d(\mu,\Sigma)$ denotes a $d$-dim Gaussian random variable with mean $\mu$ and covariance $\Sigma$. $\phi_d(x;\mu,\Sigma)$ denotes the pdf of a $N_d(\mu,\Sigma)$ random variable evaluated at $x$ and when $\Sigma$ is full rank, it is given by,
\begin{align*}
 \phi_d(x;\mu,\Sigma) = \frac{1}{(2\pi)^{d/2}\sqrt{det(\Sigma)}}\exp{(-\frac{x^T\Sigma^{-1}x}{2})}
\end{align*}
 $N_d(a, b, \mu, \Sigma)$ denotes a $d$-dim normal truncated to lie in $(a, b)$, where $a$, $b$ are $d$-dim vectors and the truncation is component-wise.
 \item $\Phi(x) = P(X \leq x)$, where $X \sim N(0, 1)$, $\Phi(a, b; \mu, \sigma^2) = P(X \in (a, b))$ when $X \sim N(\mu, \sigma^2)$. In general, $\Phi(\D; \mu, \sigma^2) = P(X \in \D)$, for $\D\in \mathcal{B}(\Re)$, where $\mathcal{B}(\Re)$ denotes the Borel $\sigma$-field over the reals. $\Phi_n(\D^n;\nu,\Delta) = P\left(X_1,\ldots,X_n \in \D^n\right)$ where $(X_1,\ldots,X_n) \sim N_n(\nu, \Delta)$ and $\D^n \in \mathcal{B}(\Re^n)$. Also for $a,b\in\Re^n$, $\Phi_n(a,b;\nu,\Delta) = P(a_i< X_i\leq b_i,1\leq i\leq n)$, where $X\sim N_n(\nu,\Delta)$. We similarly define $\Phi_n(-\infty,b;\nu,\Delta)$ and $\Phi_n(a,\infty;\nu,\Delta)$.
 \item The notion of optimality to be used throughout the paper is mean squared error optimality.
 \item Whenever we say that $\{x_n\}$ and $\{y_n\}$ have a linear state space structure, we mean the following
\begin{subequations}\label{eq: sysdef}
\begin{align}
x_{\var+1} &= Fx_{\var} + G_1w_{\var} + G_2u_\var\\
y_\var &= Hx_\var + v_\var
\end{align}
\end{subequations}
where $x_\var \in \Re^d$ is the state, $y_\var \in \Re$ is the observation, and $w_\var \in \Re^d$ and $v_\var \in \Re$ are uncorrelated Gaussian white noises with zero means and covariances $W$ and $R$, respectively. The initial state, $x_0$, of the system, is also a zero mean Gaussian with covariance $P_0$ and is uncorrelated with both $w_\var$ and $v_\var$. $u_\var$ is the control input, which is set to $0$ whenever we consider open loop estimation.
\item If $\mu$ is a measure on $\mathcal{B}\bra{\Re^k}$, then for a $\mu-$measurable function $f$, $\mu(f) \triangleq \int fd\mu$.
\item In general, $L^{p}(X,\mu)$ denotes the set of all functions defined on $X$ that are $L^p$-integrable w.r.t the measure $\mu$. Whenever, $\mu$ admits a density $p_{\mu}$, $L^p(X,\mu)$ and $L^p(X,p_{\mu})$ will be used interchangeably.
\end{enumerate}

\section{Problem Statement and Background}
\label{sec: problem}
The broader problem that one would like to solve can be cast as causal estimation of a random process $\{x_n\}$ using a quantized version, $\{q_n\}$, of the associated measurement process $\{y_n\}$. The encoding/quantization of $\{y_n\}$ into $\{q_n\}$ is determined by the information available at the encoder/observer at each time. In the classical LQG model, the observer, at each time $\var$, has access to $y_{0:\var}$, i.e., all uncoded measurements up to time $\var$ and the controller is co-located with the observer. This problem is well understood. Increasingly many modern control systems employ multiple sensors and actuators that are not co-located. Towards addressing this paradigm, there has been considerable amount of work on estimation and control under communication constraints, a representative sample being \cite{borkar, yuksel2, nair, minero, tatikonda, sahai}. A salient feature of this body of work is the presence of a single observer in the system that has access to all the measurements $\{y_n\}$ and hence these techniques do not apply in the sensor network context, for, no single sensor has access to all the uncoded measurements $\{y_n\}$.
\begin{figure}
\centering \includegraphics[scale = 0.25]{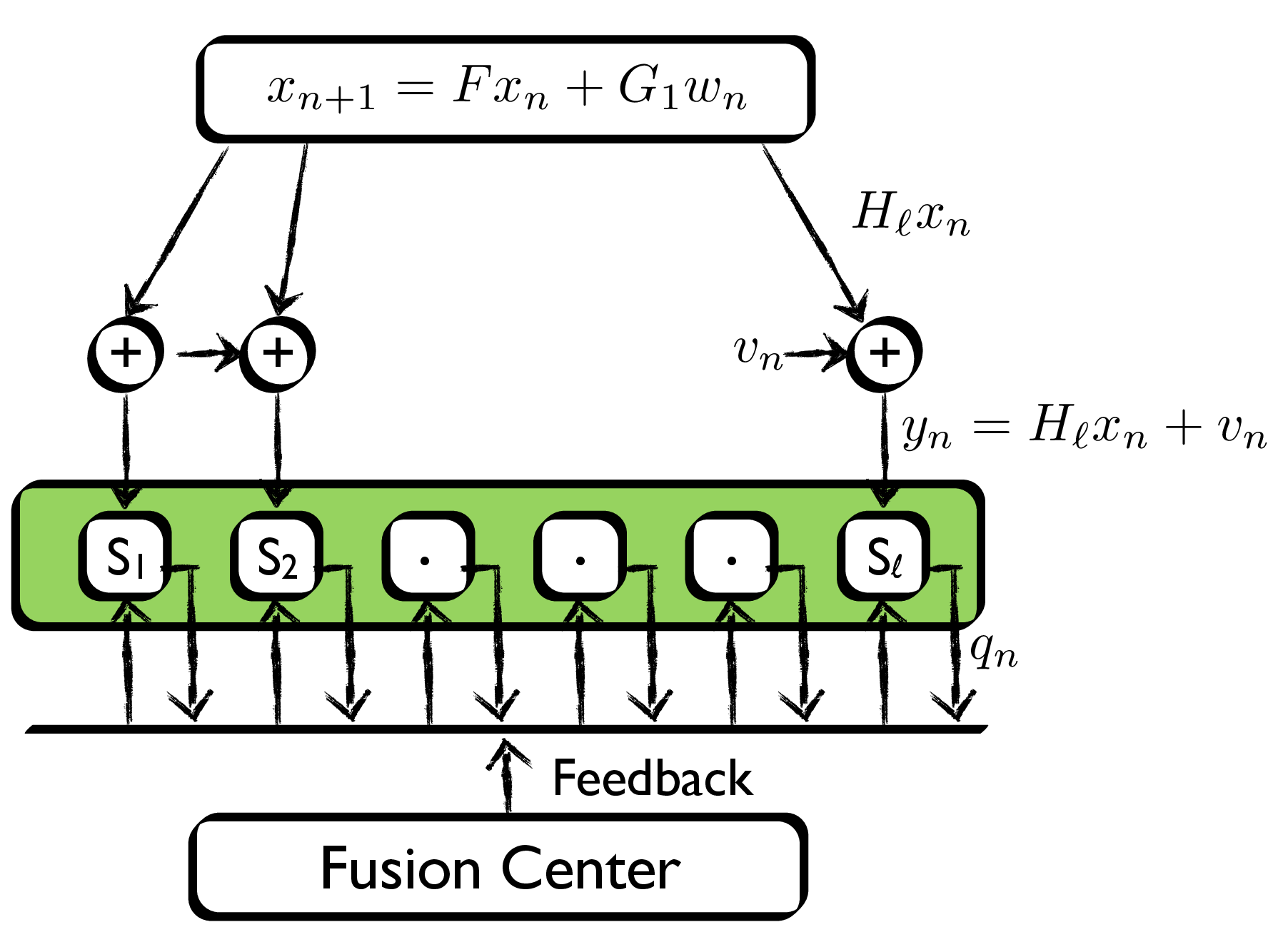}
\caption{\small WSN with a fusion center: The sensors act as data gathering devices. $S_i$ denotes the $i^{th}$ sensor and in the above figure, $S_\ell$ is making the measurement at time $n$}
\label{fig: sensor_net_cdc}
\end{figure}
In this paper, we primarily focus on the sensor network configuration in which the fusion center has sufficient power to broadcast its feedback to the sensors. Sensors are assumed to have limited power and hence their transmission of information should be limited. Here, we assume that the energy required for receiving messages is much less than that for transmitting. Fig \ref{fig: sensor_net_cdc} outlines the overall filtering paradigm\footnote{Here, we assume that the sensor communicates with the fusion center using a discrete rate-limited noiseless channel.}. So, the information available at the fusion center is $\{q_n\}$. Moreover, since the sensors do not communicate between themselves, $q_n$ is constrained to depend only on $y_n$ and $q_{0:n-1}$. For this information pattern, if the state $\{x_n\}$ is scalar and is available to the sensors, \cite{yuksel3} provides an adaptive quantization scheme that can provably track the state with an asymptotically bounded error. But when the $\{x_n\}$ is only partially observed\footnote{We assume that the measurements $\{y_n\}$ are scalar for ease of presentation but everything in this paper can be extended to the vector case}, the authors are not aware of any estimation algorithms with a provable error performance. This is an important problem and there are a number of interesting questions to be answered. In this paper, we propose an efficient particle fitering algorithm to optimally estimate the state at the fusion center using the quantized measurements $\{\Q_n\}$. Before proceeding further, we define here what we mean by a causal quantization policy. A quantizer, $\Q$, for a scalar continuous variable, is a Borel-measureable mapping from the real line to a finite set, characterized by corresponding bins $\{\mathcal{B}_i\}$ and their representations $\{a_i\}$, such that for all $i$, $\Q(x) = a_i$ iff $x\in\mathcal{B}_i$.
We use $\Qspace$ to denote an appropriate (for eg, see \cite{yuksel5}) space of such quantizers. The reason for such an informal definition is that our focus is not on finding an optimal quantization policy (for which, see \cite{yuksel5, borkar2}), its instead on computing the optimal estimator for a fixed causal quantization policy.
\begin{define}[Causal Quantization Policy]
A causal quantization policy is a sequence of quantizers $\{\Q_n\}$ such that at each time $n$, $\Q_n\in\Qspace$ is determined by the previous quantizers $\Q_{0:n-1}$ and their outputs $q_{0:n-1}$, where $q_n = \Q_n(y_n,q_{0:n-1})$.
\end{define}

An example of such a policy is \lq quantized innovations', where $q_n = \Q(y_n - \E y_n|q_{0:n-1})$ for some fixed quantizer $\Q(\cdot)$. Under the simplifying assumption that the prior $x_n|q_{0:n-1}$ is Gaussian, filtering equations of the following form have been obtained for $\hat{x}_{\var|\var} \triangleq \E x_n|q_{0:n-1}$ in \cite{SOI_KF,NTU}.
\begin{subequations}
\label{eq: mod_riccati}
\begin{align}
 &\hat{x}_{\var|\var} = \hat{x}_{n|n-1} + L\bra{q_n}\frac{P_nH^T}{\bra{HP_nH^T + R}^{1/2}}\nonumber\\
&\hat{x}_{n+1|n} = F\hat{x}_{\var|\var}\nonumber\\
&P_{\var|\var} = P_n - \lambda\frac{P_nH^THP_n}{HP_nH^T + R}\label{mod_riccati_1}\\
&P_{n+1}\triangleq P_{n+1|n} = FP_{\var|\var}F^T + G_1WG_1^T\label{mod_riccati_2}
\end{align}
\end{subequations}
The value of $\lambda$ and the mapping $L\bra{q_n}$ depend on the quantization scheme used and are detailed in \cite{NTU}. In particular, if $q_n = sign\bra{y_n - \hat{y}_{n|n-1}}$, $\lambda = \frac{2}{\pi}$ and $L\bra{q_n} = \sqrt{\frac{2}{\pi}}q_n$. Eqs \eqref{mod_riccati_1} and \eqref{mod_riccati_2} constitute the MLQ-Riccati with parameter $\lambda$. The above filter is optimal if the conditional distribution, $p\bra{x_n|q_{0:n-1}}$, is Gaussian, which we will prove is generally false. \cite{teja,KLPF} provide examples where the error performance of the filters in \cite{SOI_KF, NTU} do not track the MLQ-Riccati that they were predicted to, i.e., Eq \eqref{eq: mod_riccati}. In order to understand the problem better, we take a closer look at the conditional law of $x_n|q_{0:n}$ in the following section. 
When $\{x_n\}$ and $\{y_n\}$ are jointly Gaussian, we will provide a novel stochastic characterization of $x_n$ causally conditioned on the quantized measurement process $\{q_n\}$. This, in turn, allows us to identify the conditional density of $x_n|q_{0:n}$ to be, what we refer to as, a generalized closed skew normal distribution. We also use it to propose a novel filtering technique for the above problem which reduces to an elegant particle filter when $\{x_n\}$ and $\{y_n\}$ have  linear state space structure and outperforms the filters proposed in \cite{SOI_KF, NTU}, while providing much needed theoretical insight into the problem. 

\section{A Stochastic Characterization of the State Density Conditioned on Quantized Observations}
\label{sec: gcsn}
Suppose $\{x_n\}$ and $\{y_n\}$ are jointly Gaussian, then it is well known that the probability density of $x_n$ conditioned on $y_{0:n}$ is a Gaussian with the following parameters
\begin{align}
\label{eq: full_sto} x_n|y_{0:n} &\sim Z_n + R_{x_ny_{0:n}}R_{y_{0:n}}^{-1}y_{0:n}\quad\text{ where}\\
Z_n &\sim N_d(0,\underbrace{R_{x_n} - R_{x_ny_{0:n}}R_{y_{0:n}}^{-1}R_{y_{0:n}x_n}}_{\triangleq R^{\Delta}_{x_n,y_{0:n}}})
\end{align}
When $\{x_n\}$ has an underlying state space structure and $\{y_n\}$ is a linear measurement of $\{x_n\}$ corrupted by additive white Gaussian noise, as defined in Eq (\ref{eq: sysdef}), it is well known that the following Riccati recursion propagates the error covariance $P^{kf}_n \triangleq R^{\Delta}_{x_n,y_{0:n}} = \Vert x_n - \E x_n|y_{0:n}\Vert^2$
\begin{align}
P^{kf}_n &\triangleq P^{kf}_{n|n-1} = FP^{kf}_{\var-1|\var-1}F^T + G_1WG_1^T \nonumber\\
\label{eq: riccati_full}P^{kf}_{\var|\var} &= P^{kf}_n - \displaystyle P^{kf}_nH(HP^{kf}_nH^T + R)^{-1}H^TP^{kf}_n\\
P^{kf}_0 &= R_{x_0} \nonumber
\end{align}
The superscript \lq kf' denotes \lq Kalman filter'. We would like to address the problem of optimal estimation using a quantized version of the observation process $\{y_n\}$. Let $\{\Q_n\}$ be a causal quantization policy\footnote{The analysis goes through for more general quantization policies, for eg, $\Q_n$ can be a measurable function of the entire observation history $y_{0:n}$.}. This includes, as a special case, the method of quantizing the innovations first proposed in \cite{brockett}. We will show that the probability density of $x_n$ conditioned on the quantized measurements $q_{0:n}$ admits a characterization very similar to Eq \eqref{eq: full_sto}. We state the result in the following theorem.
\begin{thm}
 \label{thm: quant_sto}The state $x_n$ conditioned on the quantized measurements $q_{0:n}$ can be expressed as a sum of two independent random variables as follows
\begin{align}
\label{eq: quant_sto} x_n|q_{0:n} &\sim Z_n + R_{x_n,y_{0:n}}R_{y_{0:n}}^{-1}\left[y_{0:n}|q_{0:n}\right],\,\,\,\text{  where}\\
Z_n &\sim N_d(0,R^{\Delta}_{x_n,y_{0:n}})
\end{align}
\end{thm}
\begin{proof}
 The proof is fairly straightforward. The theorem will be proved by showing that the moment generating fuction of $x_n|q_{0:n}$ can be seen as the product of two moment generating functions corresponding to the two random variables in Eq \eqref{eq: quant_sto}. Note that the moment generating function of a $d-$dim random variable $X$ is given by $M_X(s) = \E e^{s^TX}$, $\forall$ $s\in\Re^{d}$. 
\begin{align*}
 p(x_n|q_{0:n}) = \int p(x_n, y_{0:n}|q_{0:n})dy_{0:n}
\end{align*}
Noting that $p(x_n| y_{0:n},q_{0:n}) = p(x_n|y_{0:n})$, we can write 
\begin{align*}
Ee^{s^Tx_n}&|q_{0:n} = \int e^{s^Tx_n}p(x_n|y_{0:n})p(y_{0:n}|q_{0:n})dx_ndy_{0:n}\\
&\stackrel{(*)}{=} e^{\frac{1}{2}s^TR^{\Delta}_{x_n,y_{0:n}} s}\underbrace{\int e^{s^TR_{x_n,y_{0:n}}R_{y_{0:n}}^{-1}y_{0:n}}p(y_{0:n}|q_{0:n})dy_{0:n}}_{\triangleq\text{ mfg of }R_{x_n,y_{0:n}}R_{y_{0:n}}^{-1}y_{0:n}|q_{0:n}}
\end{align*}
\begin{align}
\label{eq: temp1}\implies M_{x_n|q_{0:n}}(s) &= M_{Z_n}(s)M_{y_{0:n}|q_{0:n}}(R_{y_{0:n}}^{-1}R_{y_{0:n},x_n}s)
\end{align}
 where $Z_n \sim N_d(0, R^{\Delta}_{x_n,y_{0:n}})$. In getting $(*)$, we used the fact that $x_n|y_{0:n} \sim N_d(R_{x_n,y_{0:n}}R_{y_{0:n}}^{-1}y_{0:n}, R^{\Delta}_{x_n,y_{0:n}})$. For any random variable $Y$, it is easy to see that $M_Y(A^Tt) = M_{AY}(t)$. The result is now obvious from Eq \eqref{eq: temp1}. Note that if $\{x_n\}$ and $\{y_n\}$ have the state space structure, then $R^{\Delta}_{x_n,y_{0:n}} = P^{kf}_{\var|\var}$.
\end{proof}
Comparing Eqs \eqref{eq: full_sto} and \eqref{eq: quant_sto}, the only difference is the measurement vector $y_{0:n}$ being replaced by the random variable $y_{0:n}|q_{0:n}$. It is easy to see that $y_{0:n}|q_{0:n}$ is a multivariate gaussian random variable truncated to lie in the region defined by $q_{0:n}$. It is worth noting that the covariance of $x_n|q_{0:n}$, denoted by $ \Vert x_n|q_{0:n}\Vert^2 $, is given by
\begin{align*}
\Vert x_n|q_{0:n}\Vert^2 = R^{\Delta}_{x_n,y_{0:n}} + R_{x_n,y_{0:n}}R_{y_{0:n}}^{-1}\Vert y_{0:n}|q_{0:n}\Vert^2R_{y_{0:n}}^{-1}R_{y_{0:n},x_n}
\end{align*}
Stating loosely, as the quantization scheme becomes finer, $y_{0:n}|q_{0:n}$ clearly converges to $y_{0:n}$ and $x_n|q_{0:n}$ approaches a Gaussian as is well known. Using theorem \ref{thm: quant_sto}, it is easy to see that $x_n|q_{0:n}$ is not Gaussian in general, contrary to the assumption made in \cite{SOI_KF, NTU}. Infact it belongs to a class of distributions, which we call the Generalized Closed Skew Normal Distributions (GCSN) (for eg, see \cite{genton_book}), the details of which are given in the following section. Even without the gaussian assumption on the prior, we will show in the following sections that the above stochastic characterization allows an efficient numerical filtering technique whose computational complexity is comparable to the classical Kalman filter.

\subsection{The Conditional State Distribution}
Using Baye's rule, it is easy to see that 
\begin{align}
p(x_n|q_{0:n}) &= p(x_n)\frac{p(q_{0:n}|x_n)}{p(q_{0:n})}= \phi_d(x_n; 0, R_{x_n})\frac{\Phi_n(\D^n;R_{y_{0:n},x_n}R_{x_n}^{-1}x_n, R^{\Delta}_{y_{0:n},x_n})}{\Phi_n(\D;0,R_{y_{0:n}})}\label{eq: bayes_cond_density}\\
R^{\Delta}_{y_{0:n},x_n} &\triangleq R_{y_{0:n}} - R_{y_{0:n},x_n}R_{x_n}^{-1}R_{x_n,y_{0:n}}\nonumber
\end{align}
where $\D \in \mathcal{B}(\Re^n)$ is the region in which $y_{0:n}$ lies that is implied by a specific realization of the quantized measurements $q_{0:n}$. The form of the distribtion in (\ref{eq: bayes_cond_density}) is very similar to what is studied in the statistics literature as the Closed Skew Normal distribution, which is defined as follows
\begin{define}[Ch 2, \cite{genton_book}]
\label{def: CSN}
Consider $d\geq 1$, $n \geq 1$, $\mu\in\Re^d$, $\nu \in\Re^n$, $D$ an arbitrary $n\times d$ matrix, $\Sigma$ and $\Delta$ positive definite matrices of dimensions $d\times d$ and $n\times n$ respectively. Then the probability density function of the Closed Skew Normal distribution $CSN(\mu,\Sigma,D, \nu,\Delta)$ is given by 
\begin{align}
 \label{eq: CSN_pdf}
CSN(y;\mu,\Sigma,D,\nu,\Delta) = \phi_d(y;\mu,\Sigma)\frac{\Phi_n(-\infty, D(y-\mu);\nu,\Delta)}{\Phi_n(-\infty, 0;\nu,\Delta+D\Sigma D^T)}
\end{align}
\end{define}
Stochastically, $CSN(\mu,\Sigma,D, \nu,\Delta)$ is the probability density of $X|Z - D(X-\mu) < 0$, where $X\sim N_d(\mu, \Sigma)$ and $Z\sim N_n(\nu,\Delta)$ are independent and the inequality $Z - D(X-\mu) < 0$ is component-wise. One can arrive at this characterization by a simple application of the Baye's rule. Skew elliptical distributions generated a lot of interest (\cite{azzalini_multivariate_1996,valle_fundamental_2005, genton_book, genton_moments, genton_discussion_2005, skewed_kalman}) because they provide a much needed tool to handle skewness in statistical modeling and have a good number of properties in common with the standard normal distribution, such as closure under marginlization and conditioning. In particular, such skew distributions arise via hidden truncation processes. In the context of estimation using quantized measurements, this truncation is the consequence of quantization, so such skew distributions naturally show up here. For example, consider the sign of innovation scheme given by $q_n = sign(y_n - \hat{y}_{n|n-1})$, where $\hat{y}_{n|n-1} = \E y_n|q_{0:n-1}$. In this setup, as will be shown below, the conditional law of $x_n|q_{0:n}$ is a Closed Skew Normal distribution. Consider a fixed observation record $q_{0:n}$, then we have
\begin{align}
\label{eq: soi1}
 p(q_{0:n}) &= Pr\bigl(q_i(y_i - \hat{y}_{i|i-1}) \geq 0,\forall\,\, 0\leq i\leq n\bigr) \nonumber\\
&= \idotsint\limits_{\substack{q_iy_i \geq q_i\hat{y}_{i|i-1}\\0\leq i\leq n}} \phi_{n+1}\left(y_{0:n};0,R_{y_{0:n}}\right)dy_{0:n}\nonumber\\
&= \idotsint\limits_{\substack{\xi_i \geq q_i\hat{y}_{i|i-1}\\0\leq i\leq n}} \phi_{n+1}\left(\xi_{0:n};0,R_{\xi_{0:n}}\right)d\xi_{0:n},\,\,\,\text{where}\,\,\,\mathbf{\xi}_i = q_iy_i,\,\,R_{\xi_{0:n}} = \text{diag}(q_{0:n})R_{y_{0:n}}\text{diag}(q_{0:n})\nonumber\\
&= \Phi_{n+1}\left(-\infty,0;\mathbf{\nu}_n,R_{\xi_{0:n}}\right),\,\,\,\text{where}\,\,\,\mathbf{\nu}_n  = [q_0\hat{y}_{0|-1},\ldots,q_n\hat{y}_{n|n-1}]^T
\end{align}
Similarly, one can show that 
\begin{align}
\label{eq: soi2}
 p(q_{0:n}|x_n) = \Phi_{n+1}\left(-\infty,R_{\xi_{0:n},x_n}R_{x_n}^{-1}x_n;\mathbf{\nu}_n,R^{\Delta}_{\xi_{0:n},x_n}\right)
\end{align}
where $R_{\xi_{0:n},x_n} = \text{diag}(q_{0:n})R_{y_{0:n},x_n}$ and $R^{\Delta}_{\xi_{0:n},x_n} = R_{\xi_{0:n}} - R_{\xi_{0:n},x_n}R_{x_n}^{-1}R_{x_n,\xi_{0:n}}$. Using eqs (\ref{eq: soi1}) and (\ref{eq: soi2}), we get
\begin{align}
 p(x_n|q_{0:n}) &= p(x_n)\displaystyle\frac{p(q_{0:n}|x_n)}{p(q_{0:n})}\nonumber\\
&= \phi_d\left(x_n;0,R_{x_n}\right)\displaystyle\frac{\Phi_{n+1}\left(-\infty,R_{\xi_{0:n},x_n}R_{x_n}^{-1}x_n;\mathbf{\nu}_n,R^{\Delta}_{\xi_{0:n},x_n}\right)}{\Phi_{n+1}\left(-\infty,0;\mathbf{\nu}_n,R_{\xi_{0:n}}\right)}\nonumber\\
\implies p(x_n|q_{0:n}) &= CSN\left(x_n;0,R_{x_n},R_{\xi_{0:n},x_n}R_{x_n}^{-1},\mathbf{\nu}_n,R^{\Delta}_{\xi_{0:n},x_n}\right)
\end{align}

In order to capture the effect of a general quantization scheme, one would need a straightforward generalization of the CSN distirbution. It is obtained by considering the probability density of \\
$X|\left(Z - D(X-\mu) \in \D^n\right)$, where $\D^n\in\mathcal{B}(\Re^n)$. This will result in probability density functions of the form (\ref{eq: bayes_cond_density}).  We will refer to such distributions as the Generalized Closed Skew Normal Distributions (GCSN), which are formally defined as follows.
\begin{define}
\label{def: GCSN}
 For $x\,\in\, \Re^n$ and $\D^n\in\mathcal{B}(\Re^n)$, we define the generalized closed skew-normal distribution, \\$\GCSNx$, as follows
\begin{align}
\label{eq: def_GCSN}
&\GCSNx \triangleq\  \phi_d\bra{x;\mu, \Sigma}L_{d,n}(.)\nonumber\\
&L_{d,n}(.) = \frac{\Phi_n\bra{\D^n;D\bra{x-\mu},\Delta}}{\Phi_n\bra{\D^n;0,\Delta + D\Sigma D^T}}
\end{align}
\end{define}
Now, suppose $\{x_n\}$ and $\{y_n\}$ have the state space structure of (\ref{eq: sysdef}) and suppose $W$ is positive definite for all $n\geq 0$. Then the evolution of the conditional state distribution with time is completely characterized by the following theorem.

\begin{thm}[Conditional State Distribution]
  \label{thm: state_density}
The probability density function of $x_n|q_{0:n}$ is given by $GCSN_{d,n+1}\left(x_n; 0, R_{x_n}, R_{y_{0:n},x_n}R_{x_n}^{-1}, \D^{n+1}, R^{\Delta}_{y_{0:n},x_n}\right)$. The recursions relating the parameters of the distributions of $x_{n-1}|q_{0:n-1}$ and $x_n|q_{0:n}$ are given by
\begin{subequations}
\label{eq: kalman_filter_quant}
\begin{align}
R_{x_n} &= FR_{x_{n-1}}F^T + G_1WG_1^T,\,\,R_{y_{0:n},x_n} = \begin{bmatrix}R_{y_{0:n-1},x_n}A^T\\ H\end{bmatrix}\\
R_{y_{0:n}} &= \begin{bmatrix} R_{y_{0:n-1}} & R_{y_{0:n-1},x_n}A^TH^T\\ HAR_{x_n,y_{0:n-1}} & R + HR_{x_n}H^T\end{bmatrix},\,\,\,R_{y_0} = R + HR_{x_0}H^T\\ 
\D^{n+1} &= \D^n\cap \{y_{0:n}\in\Re^n| Q^{_{\var}}(y_{0:n})=q_n\}
\end{align}
\end{subequations}
\end{thm}
\begin{proof}
See Appendix.
\end{proof}
When the full measurements $y_{0:n}$ are available, the conditional state density is completely characterized by its mean and covariance which are propagated by the traditional Kalman filtering equations (Ch 9, \cite{Linear_Estimation}). When only the quantized measurements are available, it is interesting to note that the conditional state distribution is completely characterized by a finite number of parameters which are propogated as given in theorem \ref{thm: state_density}. So, Eq (\ref{eq: kalman_filter_quant})  constitutes the equivalent of the traditional Kalman filtering equations in the case when only the quantized measurements are available. In fact, one can write non-trivial formulae for the mean and covariance of a GCSN, but computing them will quickly become infeasible since the dimensions of some of the matrices involved in Eq (\ref{eq: kalman_filter_quant}) grow with time. Except, $\D^{n}$, all other parameters are independent of the specific realization of the quantized measurements and hence, in principle, can be propagated offline. Theorem \ref{thm: quant_sto} can be used to translate any results on the properties of the closed skew normal distribution into additional insights on the current problem. Next we discuss a special case where we derive closed form Kalman-like recursions for the mmse estimate of the state and the corresponding estimation error. 

\subsection{A Comment on Quantizing the true innovation}
Suppose $\{x_n\}$ and $\{y_n\}$ have the linear state space structure of (\ref{eq: sysdef}) with $\{y_n\}$ being a scalar measurement process. The innovations process associated to $\{y_n\}$ is denoted by $\{e_n\}$, i.e., $e_n = y_n - \E y_n|\mathbf{y}_{n-1}$ and $R_{e_n}\triangleq \Vert e_n\Vert^2$. The following notation shall be used in the rest of the paper.
\begin{align*}
 \hat{x}_{n|m} &\triangleq Ex_n|q_{0:m},\,\, \hat{x}_n \triangleq \hat{x}_{n|n-1},\,\, \hat{x}^{kf}_{n|m} = Ex_n|y_{0:m},\,\,\hat{x}^{kf}_n\triangleq \hat{x}^{kf}_{n|n-1}\\
P_{n|m} &\triangleq \Vert x_n - \hat{x}_{n|m}\Vert^2,\,\, P_n\triangleq P_{n|n-1}\\
P^{kf}_{n|m} &\triangleq  \Vert x_n - \hat{x}^{kf}_{n|m}\Vert^2,\,\, P^{kf}_n\triangleq P^{kf}_{n|n-1}
\end{align*}
For ease of exposition, we assume a fixed quantizer $\Q(.)$ whose quantization intervals are given by $\{(z_0,z_1),(z_1,z_2),\ldots,(z_{\ell-1},z_\ell),(z_\ell,z_{\ell+1})\}$, where $z_0 = -\infty$ and $z_{\ell+1} = \infty$. So, if $q_n = \Q\bigl(e_n/R^{1/2}_{e_n}\bigr)$, then a realization of $q_{0:n}$ would imply that $e_j/R^{1/2}_{e_j} \in (z_{l_j},z_{l_j+1}) ,\,\,j\leq n$ for some $0\leq l_j\leq\ell$. With this setup, we have the following result. 
\begin{thm}[Optimal Estimation Using Quantized 'true' Innovations]
\label{thm: true_innov}
The mmse estimate of $x_n$ using $q_{0:n}$, denoted by $\hat{x}_{\var|\var}$, and the associated estimation error, denoted by $P_{\var|\var}$, are given recursively by the following equations
\begin{subequations}
\label{eq: quant_true_inno}
 \begin{align}
	\hat{x}_{\var|\var} &= A\hat{x}_{\var-1|\var-1} + \displaystyle\frac{P^{kf}_nH^T}{\sqrt{HP^{kf}_nH^T + R}}\frac{\phi(z_{l_n}) - \phi(z_{l_n+1})}{\Phi(z_{l_n+1}) - \Phi(z_{l_n})}\\
	P_{\var|\var} &= AP_{\var-1|\var-1}A^T - \displaystyle\alpha\frac{P^{kf}_nH^THP^{kf}_n}{HP^{kf}_nH^T + R} + G_1WG_1^T\label{eq: true_innov_recur}\\
	\alpha &= \sum_{k=0}^{\ell }\frac{(\phi(z_{k}) - \phi(z_{k+1}))^2}{\Phi(z_{k+1}) - \Phi(z_{k})},\,\,\,z_{\ell+1}\triangleq \infty,\,\,\,z_0\triangleq -\infty\\
	P^{kf}_{n+1} &= AP^{kf}_nA^T - \frac{AP^{kf}_nH^THP^{kf}_nA^T}{HP^{kf}_nH^T + R}+ G_1WG_1^T
 \end{align}
\end{subequations}
\end{thm}
\begin{proof}
 See the Appendix.
\end{proof}
\begin{cor}[Convergence of the Error Covariance]
\label{cor: conv_trueinnov}
 Suppose $F$ is stable and $\Lambda$ is the unique positive semi-definite solution to the discrete-time Lyapunov equation
\begin{align*}
 \Lambda = F\Lambda F^T + G_1WG_1^T
\end{align*}
 and let $P^{kf}$ be the unique positive semi-definite solution to the following discrete-time algebraic Riccati equation (DARE)
\begin{align*}
Z = FZF^T - \displaystyle\frac{FZH^THZF^T}{HZH^T + R} + G_1WG_1^T
\end{align*}
And let $P^f = P^{kf} - \displaystyle\frac{P^{kf}H^THP^{kf}}{HP^{kf}H^T + R}$. Then the error covariance $P_{\var|\var} \longrightarrow P$, where $P$ is given by
\begin{align}
 \label{eq: limit_true_inno}
P = \alpha P^f + (1-\alpha)\Lambda
\end{align}
Further, if $F$ is unstable, then, irrespective of the quantization scheme used, $P_{\var|\var}\longrightarrow\infty$.
\end{cor}
\begin{proof}
 See the Appendix.
\end{proof}

For a fixed number of quantization levels, the value of $\alpha$ can be optimized by choosing $\{z_j\}_{j=1}^{\ell}$ appropriately. The above innovation coding scheme was introduced in \cite{borkar} but closed form expressions for the optimal state estimate and the corresponding estimation error of the form stated above were not presented. The fact that $P_{\var|\var}$ diverges if $F$ is unstable seems to be common knowledge (for eg, see \cite{tatikonda2}), the authors are not aware of a concrete proof before this work.

Note that the above scheme is not suited for distributed applications where no observer in the network has enough information to compute the innovations process. In general, the problem of optimal state estimation using quantized measurements does not admit an analytically tractable solution like the one above. 
This necessitates a numerical solution. But, using the insight of Theorem \ref{thm: quant_sto}, we will show that $\hat{x}_{\var|\var}$ can be numerically approximated with a complexity that is, in most cases, comparable to the classical Kalman filter. In the following section, we outline the general particle filtering technique which will then be specialized to solve the problem of optimal state estimation using quantized measurements by exploiting Theorem \ref{thm: quant_sto}. 

\section{Application to Particle Filtering}
A promising approach to recursive estimation in non-linear problems is particle filtering. For easy reference, a basic bootstrap filter for the case when $\{x_n\}$ and $\{y_n\}$ have state space structure of (\ref{eq: sysdef}) is outlined below. Let $\{\alphaN\}_{M\geq1}$ be a sequence of positive integers with a limit $\alpha$, which could be infinity. 
\label{sec: the_filter}
\begin{flushleft}
 \begin{tabular}{p{160mm}}
\hline
 Alg 1. Particle Filter\\
\hline
\end{tabular}
\begin{enumerate}
 \item Set $\var=0$. For $i = 1,\cdots,M\alphaN$, initialize the particles, $x^{i}_{0|-1} \sim p(x_0)$ and set $\hat{x}_{0|-1} = 0$
 \item At time $\var$, using measurement $\textstyle q_{\var} = \Q_{\var}\bra{y_{0:n}}$, the importance weights are calculated as follows
\begin{align*}
w^{i}_{\var} = p\bra{q_n|x_n = x^{i}_{n|n-1},q_{0:n-1}}.
\end{align*}
 \item Measurement update is given by
\begin{align*}
 \hat{x}^{pf,M}_{\var|\var} = \sum_{i=1}^{M\alphaN}\overline{w}^{i}_{\var}x^{i}_{\var|\var-1}
\end{align*}
where $\overline{w}^i_n$ are the normalized weights, i.e., 
\begin{align*}
 \overline{w}^{j}_{\var} = \frac{w^{j}_{\var}}{\sum_{i=1}^Mw^{i}_{\var}}
\end{align*}
 \item Resample $M$ particles from the above $M\alphaN$ particles with replacement as follows. Generate i.i.d random variables $\{J_\ell\}_{\ell = 1}^M$, such that $P(J_\ell = i) = \overline{w}^i_n$. Then
\begin{align*}
x^{\ell}_{\var|\var} = x^{J_\ell}_{\var}
\end{align*}
 \item For $i=1,\cdots,M\alphaN$, predict new particles according to,
\begin{align*}
 x^{j}_{\var+1|\var} &\sim p\bra{x_{\var + 1}|x_n = x^{i}_{\var|\var}},\text{ i.e.,}\\
x^j_{n+1|n} &= Fx^i_{\var|\var} + G_1w^j_{\var},\,\, (i-1)\alphaN+1\leq j\leq i\alphaN
\end{align*}
where $\{w^j_{\var}\}_{j=1}^{M\alphaN}$ are i.i.d $\phi_d(0,W)$
\item Set $\hat{x}^{pf,M}_{\var+1|\var} = F\hat{x}^{pf,M}_{\var|\var}$. Also, set $\var = \var+1$ and iterate from step 2.
\end{enumerate}
\begin{tabular}{p{160mm}}
\hline
\\
\end{tabular}
\end{flushleft}
For example, if one uses the sign of innovation scheme, $q_n = sign(y_n - \hat{y}_{n|n-1})$, it is easy to see that the importance weights are given by $w^i_{\var} = \Phi\bra{q_nH(x^i_{n|n-1}-\hat{x}_{n|n-1});0,R}$.
The particles in Alg 1 describe the conditional state density $p\bra{x_n|q_{0:n}}$ and simulations suggest that one needs upwards of a thousand particles to get satisfactory error performance for most systems. In what follows, we use Theorem \ref{thm: quant_sto} to develop a novel particle filtering technique (KLPF) which converges to the optimal filter much faster than the generic filter outlined in Alg 1. The difference lies in using particles to describe a probability density with a much smaller covariance than the conditional state density. We begin by noting that
\begin{align}
 \label{eq: truncated_mean}\E x_n|q_{0:n} = R_{x_n,y_{0:n}}R_{y_{0:n}}^{-1}\E y_{0:n}|q_{0:n}
\end{align}
So, it should suffice to propogate particles drawn from $R_{x_n,y_{0:n}}R_{y_{0:n}}^{-1}y_{0:n}|q_{0:n}$. For notational convenience, define 
\begin{align}
\label{eq: kalman_estimate}
 \xi_{\var} = R_{x_n,y_{0:n}}R_{y_{0:n}}^{-1}y_{0:n}
\end{align}
The Kalman Like Particle filter propagates the conditional law $\xi_{\var}|q_{0:n}$. Note that $\hat{x}_{\var|\var} = \\E\xi_{\var}|q_{0:n}$.

Under a causal quantization policy, the quantizer output, $q_n$ at time $n$, is obtained by quantizing a scalar valued function of $y_n,q_{0:n-1}$. So, upon receiving $q_n$ and using the previously received quantized values $q_{0:n-1}$, the fusion center infers that $y_n \in \mathcal{S}(q_{0:n})$ for some Borel measurable set $\mathcal{S}(q_{0:n})$. We write the interval as $\mathcal{S}(q_{0:n})$ to emphasize that everything is conditioned on a fixed observation record $q_{0:n}$. 

Inorder to develop a particle filter to propogate $\xi_{\var}|q_{0:n}$, one needs to compute the likelihood ratio $p(\xi_{\var-1}|q_{0:n})/p(\xi_{\var-1}|q_{0:n-1})$ and the transition from $\xi_{\var-1}$ to $\xi_{\var}$.
\begin{lem}
The likelihood ratio between the conditional laws $\xi_{\var-1}|q_{0:n}$ and $\xi_{\var-1}|q_{0:n-1}$ is given by
 \begin{align}
  \label{eq: likelihood}
	\displaystyle\frac{p(\xi_{\var-1}|q_{0:n})}{p(\xi_{\var-1}|q_{0:n-1})} &\propto \Phi\bra{\mathcal{S}(q_{0:n});HF\xi_{\var-1},R_{e_n}}
 \end{align}
\end{lem}
\begin{proof}
 An application of Baye's rule gives
 \begin{align*}
  \displaystyle\frac{p(\xi_{\var-1}|q_{0:n})}{p(\xi_{\var-1}|q_{0:n-1})} = \displaystyle\frac{P(q_n|q_{0:n-1},\xi_{\var-1})}{P(q_n|q_{0:n-1})} \propto P(q_n|q_{0:n-1},\xi_{\var-1})
 \end{align*}
Now, we have
\begin{align*}
 P(q_n|q_{0:n-1},\xi_{\var-1}) &= E\left[\bra{\mathbb{I}_{y_n\in\mathcal{S}(q_{0:n})}}|q_{0:n-1},\xi_{\var-1}\right]\\
 &= E\left[E\bra{\mathbb{I}_{y_n\in\mathcal{S}(q_{0:n})}}|\mathbf{y}_{n-1}\right]|q_{0:n-1},\xi_{\var-1} \\
 &= E\,\Phi\bra{\mathcal{S}(q_{0:n});HF\xi_{\var-1},R_{e_n}}|q_{0:n-1},\xi_{\var-1}\\
 &= \Phi\bra{\mathcal{S}(q_{0:n});HF\xi_{\var-1},R_{e_n}}
\end{align*}
\end{proof}
Now, observe that $\xi_{\var}$ is the mmse estimate of the state $x_n$ given $y_{0:n}$. Since $\{x_n\}$ and $\{y_n\}$ have the state space structure, it is well known that the Kalman filter propagates $\xi_{\var}$ recursively as follows
\begin{subequations}\label{eq: transition}
 \begin{align}
  \xi_{\var} &= F\xi_{\var-1} + K^{f}_n\bigl(y_n - HF\xi_{\var-1}\bigr),\,\,\text{where}\\
K^{f}_n&=\displaystyle\frac{P^{kf}_nH^T}{HP^{kf}_nH^T + R}
 \end{align}
\end{subequations}
Eq \eqref{eq: transition} completely describes the transition from $\xi_{\var-1}$ to $\xi_{\var}$. From the particle filtering perspective, if $\{\xi^i_{\var-1|\var}\}$ are samples from $\xi_{\var-1}|q_{0:n}$, then one can generate samples $\{\xi^i_{\var|\var}\}$ from $\xi_{\var}|q_{0:n}$ by first observing that 
\begin{align*}
 p(\xi_n|q_{0:n}) &= \int p(\xi_n,\xi_{n-1}|q_{0:n})d\xi_{n-1}\\
		  &= \int p(\xi_n|\xi_{n-1},q_{0:n})p(\xi_{n-1}|q_{0:n})d\xi_{n-1}\\
		  &= \int p\left(K^f_n y_n = \xi_n - F\xi_{n-1} + K^f_nHF\xi_{n-1}|\xi_{n-1},q_{0:n}\right)p(\xi_{n-1}|q_{0:n})d\xi_{n-1}
\end{align*}
This suggests that, for each $\xi^i_{n-1|n}$, we can obtain a sample $\xi^i_{n|n}$ from $\xi_n|q_{0:n}$ by first generating
 $y^i_{\var|\var}$ from $y_n|\xi^i_{n-1|n},q_{0:n}$ and then setting
\begin{align}
\label{eq: KLPF1}
  \xi^i_{\var|\var} &= F\xi^i_{\var-1|\var} + K^{f}_n\bigl(y^i_{\var|\var} - HF\xi^i_{\var-1|\var}\bigr)
\end{align}
It is easy to see that $y_n|\bra{\xi_{\var-1},q_{0:n}} \sim N\bra{\mathcal{S}(q_{0:n});HF\xi_{\var-1},R_{e_n}}$.
We can now desribe the Kalman Like Particle filter as follows.
\begin{flushleft}
 \begin{tabular}{p{160mm}}
\hline
 Alg 3. Kalman Like Particle Filter (KLPF)\\
\hline
\end{tabular}
\begin{enumerate}
 \item At $n=0$, generate $\{y^i_{0|0}\}_{i=1}^{M\alphaN} \sim N(\mathcal{S}(q_0); 0, R_{y_0})$. Compute $\xi^i_{0|0} = K^f_{0}y^i_{0|0}$
 \item At time $\var$, for each particle $\{\xi^i_{\var-1|\var-1}\}$, compute the weight as 
\begin{align}
w^i_{\var} = \Phi\bra{\mathcal{S}(q_{0:n});HF\xi^i_{\var-1|\var-1}, R_{e_n}}
\end{align}
Normalize the weights to get $\overline{w}^i_{\var} = \frac{w^i_{\var}}{\sum_{i=1}^{M\alphaN}w^i_{\var}}$
\item Resample $M$ particles from the above $M\alphaN$ particles with replacement as follows. Generate i.i.d random variables $\{J_\ell\}_{\ell = 1}^M$, such that $P(J_\ell = i) = \overline{w}^i_n$. Then
\begin{align*}
\xi^\ell_{\var-1|\var} = \xi^{J_{\ell}}_{\var-1|\var-1}
\end{align*}
 \item Measurement update: Generate $y^i_{\var|\var}$ i.i.d from $\phi\bra{\mathcal{S}(q_{0:n});HF\xi^{\ell}_{\var-1|\var}, R_{e_n}}$, for $(\ell-1)\alphaN+1 \leq i \leq \ell\alphaN$
 and obtain the new particles $\{\xi^i_{\var|\var}\}$ as follows
\begin{align}
 \xi^i_{\var|\var} = F\xi^\ell_{\var-1|\var} + K^{f}_n\left(y^i_{\var|\var} - HF\xi^\ell_{\var-1|\var}\right)
\end{align}
The measurement updated estimate is given by $\xklpf_{\var|\var} = \frac{1}{M\alphaN}\sum_{i=1}^{M\alphaN}\xi^i_{\var|\var}$
\item Set $\xklpf_{\var+1|\var} = F\xklpf_{\var|\var}$. Also, set $\var = \var+1$ and iterate from step 2.
\end{enumerate}
\end{flushleft}


\section{Consistency and Convergence of the KLPF}
There is a vast body of literature on the convergence behavior of particle filters, \cite{chrisan, IIHMM, vanHandel, budhiraja} being a representative sample. In this section, we will show that $M^{1/2}\bra{\xklpf_{\var|\var} - \hat{x}_{\var|\var}}$ and \\$M^{1/2}\bra{\hat{x}^{pf,M}_{\var|\var} - \hat{x}_{\var|\var}}$ converge in distribution to zero mean Gaussian random variables. In particular, the former converges to a Gaussian random variable with a much smaller variance than the latter. For ease of exposition, we present all results for a scalar valued state space model, i.e., $x_n \in \Re$. This can be extended to the vector case by treating $x_n$ one component at a time and is straightforward. Most of the literature on the convergence of particle filters assumes the traditional measurement model, where the current measurement, conditioned on the current state, is independent of the past measurements. This is clearly not true for the quantization scheme we are considering. $q_n$ is not undependent of $q_{0:n-1}$ conditioned on $x_n$. But the techniques themselves are quite general and can be easily extended to the more general measurement model at hand. Before presenting the convergence results on the particle filters proposed in the previous section, we need to introduce a couple of simple definitions. A sample of particles $\{z^i\}_{i=1}^M$ with associated weights $\{w^i\}_{i=1}^M$ is said to constitute a weighted sample $\{z^i,w^i\}_{i=1}^M$. For such a sample, consistency and asymptotic normality are defined as follows. 
\begin{define}[Consistency]
 The weighted sample $\{(z^i, w^i)\}_{1\leq i\leq M}$ is said to be consistent for the probability measure $\nu$ and the set $\mathcal{C} \subseteq L^1\bra{\Re, \nu}$ if for any $f\in\mathcal{C}$, 
\begin{align*}
 \sum_{i=1}^M\displaystyle\frac{w^i}{\sum_{j=1}^M w^j}f\bra{z^i} \stackrel{P}{\longrightarrow} \nu(f),\quad as\,\,\,M\rightarrow \infty
\end{align*}

\end{define}

\begin{define}[Asymptotic Normality]
 Let $\F$ be a class of real-valued measurable functions on $\Re$, let $\sig$ be a non-negative function on $\F$, and let $\{\alpha_M\}$ be a non-decreasing real sequence diverging to infinity. We say that the weighted sample $\{(z^i, w^i)\}_{1\leq i\leq M}$ is asymptotically normal for $(\nu, \F, \sig, \{\alpha_M\})$ if for any function $f\in\F$, it holds that $\nu(|f|) < \infty$, $\sig^2(f) < \infty$ and
\begin{align}
 \alpha_M\sum_{i=1}^M\displaystyle \frac{w^i}{\sum_{j=1}^M w^j}\left[f(z^i) - \nu(f)\right] \stackrel{\mathcal{D}}{\longrightarrow} N\bra{0,\sig^2(f)},\quad\text{as }\quad M\longrightarrow \infty
\end{align}
\end{define}
In words, asymptotic normality implies that the estimation error is distributed as a zero-mean Gaussian with a fixed variance that is independent of the number of samples, $M$, when $M$ is large. Note that consistency follows from asymptotic normality.

We present the convergence results for the case $\alphaN \rightarrow \infty$ since it allows a clean interpretation of the asymptotics. These can be extended to the more general case of $\alphaN \rightarrow \alpha > 0$ at the expense of more involved notation without giving any additional insight into the problem. Also, if a measure $\nu$ admits a density $p$, we use $\nu$ and $p$ interchangeably and the context would make it clear. 
\begin{thm}[Weak convergence of Alg 1]
\label{thm: alg1conv}
The following holds true
\begin{enumerate}
 \item If $\{x^i_{0|-1},1\}_{i=1}^{M\alphaN}$ is consistent for $\Bigl(p(x_{0}), L^1\bigl(\Re, p(x_{0})\bigr)\Bigr)$, then for any $\var > 0$, $\{x^i_{\var|\var}\}_{i=1}^{M}$ is consistent for $\Bigl(p\bra{x_{\var}|\mathbf{q}_{\var}}, L^1\bigl(\Re, p\bra{x_{\var}|\mathbf{q}_{\var}}\bigr)\Bigr)$
 \item If in addition $\{x^i(0|-1),1\}_{i=1}^{M\alphaN}$ is asymptotically normal for \\
$\Bigl(p(x_{0}), L^2\bigl(\Re,p(x_{0})\bigr),\text{Var}_{p(x_{0})}(.), \sqrt{M\alphaN}\Bigr)$, then for any $\var > 0$, $\{x^i_{\var|\var}\}_{i=1}^{M}$ is asymptotically normal for $\Bigl(p\bra{x_{\var}|\mathbf{q}_{\var}}, L^2\bigl(\Re,p\bra{x_{\var}|\mathbf{q}_{\var}}\bigr),\text{Var}_{p\bra{x_{\var}|\mathbf{q}_{\var}}}(.), \sqrt{M}\Bigr)$, in particular
\begin{align}
 \sqrt{M}\bra{\hat{x}^{pf,M}_{\var|\var} - \hat{x}_{\var|\var}} \stackrel{\mathcal{D}}{\longrightarrow} N\bra{0, \Vert x_n - \hat{x}_{\var|\var}\Vert^2}
\end{align}
\end{enumerate}
\end{thm}
In particular, whenever $\lim\sup_n \Vert x_n - \hat{x}_{\var|\var}\Vert^2 < \infty$, the above result implies that $\hat{x}^{pf,M}_{\var|\var} \rightarrow \hat{x}_{\var|\var}$ as $M \rightarrow \infty$. 
\begin{thm}[Weak Convergence of Alg 2]
\label{thm: alg2conv}
The following holds true
\begin{enumerate}
 \item If $\{\xi^i_{0|0},1\}_{i=1}^{M\alphaN}$ is consistent for $\Bigl(p(\xi_{0|0}), L^1\bigl(\Re, p(\xi_{0|0})\bigr)\Bigr)$, then for any $\var > 0$, $\{\xi^i_{\var|\var}\}_{i=1}^{M}$ is consistent for $\Bigl(p\bra{\xi_{\var}|\mathbf{q}_{\var}}, L^1\bigl(\Re, p\bra{\xi_{\var}|\mathbf{q}_{\var}}\bigr)\Bigr)$
 \item If in addition $\{\xi^i_{0|0},1\}_{i=1}^{M\alphaN}$ is asymptotically normal for \\
$\Bigl(p(\xi_{0|0}), L^2\bigl(\Re,p(\xi_{0|0})\bigr),\text{Var}_{p(\xi_{0|0})}(.), \sqrt{M\alphaN}\Bigr)$, then for any $\var > 0$, $\{\xi^i_{\var|\var}\}_{i=1}^{M\alphaN}$ is asymptotically normal for $\Bigl(p\bra{\xi_{\var}|\mathbf{q}_{\var}}, L^2\bigl(\Re,p\bra{\xi_{\var}|\mathbf{q}_{\var}}\bigr),\sigma_{\var|\var}, \sqrt{M\alphaN}\Bigr)$, in particular, for $f(x)=x$,
 \begin{align}
  \sqrt{M}\bra{\xklpf_{\var|\var} - \hat{x}_{\var|\var}} &\stackrel{\mathcal{D}}{\longrightarrow} N(0,\sigma_{\var|\var}^2(f)),\text{ where}\\
 \sigma_{\var|\var}^2(f) &\leq \Vert \xi_{\var} - \hat{\xi}_{\var|\var}\Vert^2 = R_{x_n,y_{0:n}}R_{y_{0:n}}^{-1}\Vert y_{0:n}|q_{0:n}\Vert^2 R_{y_{0:n}}^{-1}R_{y_{0:n},x_n}
 \end{align}
\end{enumerate}
\end{thm}
Proofs for Theorem \ref{thm: alg1conv} and Theorem \ref{thm: alg2conv} follow from a straightforward extension of the results in chapter 9 of \cite{IIHMM}. Now, note that the asymptotic normality and consistency of $\{\xi^i_{0|0}\}$ and $\{x^i_{0|-1}\}$ follows from the fact that they are drawn i.i.d from $p(\xi_{0}|q_0)$ and $p(x_0)$ respectively. This observation coupled with Theorem \ref{thm: alg1conv} and Theorem \ref{thm: alg2conv} proves the correctness of the brute force particlef filter and the KLPF. In addition to proving the correctness of the KLPF, Theorem \ref{thm: alg2conv} proves that the asymptotic variance of the estimates from Alg 2 is typically much smaller than that for Alg 1. The particles in the KLPF describe the random variable $R_{x_n,y_{0:n}}R_{y_{0:n}}^{-1}y_{0:n}|q_{0:n}$. Its variance decreases to zero as the number of quantization levels increases. On the other hand, the variance of $x_n|q_{0:n}$ cannot be smaller than $P^{kf}_{n|n}$. As a result KLPF needs dramatically fewer particles as the quantization becomes finer. This will be demonstrated through examples in Section \ref{sec: sims}. In practice, for most systems, $\Vert \xi_{\var} - \hat{\xi}_{\var|\var}\Vert^2$ is much smaller than $\Vert x_n - \hat{x}_{\var|\var}\Vert^2$. In such examples, simulations suggest that the KLPF delivers close to optimal performance, i.e., $\left| \hat{x}^{klpf,M}_{n|n} - \hat{x}_{n|n}\right|$ is small with high probability, for $M \leq 100$. Though Theorems \ref{thm: alg1conv} and \ref{thm: alg2conv} prove the correctness and characterize the asymptotic behavior of the particle filters, there is more to be understood about the rates of convergence of the two algorithms. That is, in practice one would be interested in bounding $P\left(|\hat{x}^{klpf,M}_{n|n} - \hat{x}_{n|n}| > B\right)$ for finite $M$. All such results in the existing literature (for e.g., \cite{chrisan}) are available only for bounded functions of the state. Clearly functions of the form $f(x) = x$, which is what we are interested in, are not bounded. Note that asymptotic normality only tells us that 
\begin{align*}
 P\left(\sqrt{M}|\hat{x}^{klpf,M}_{n|n} - \hat{x}_{n|n}| > B\right) \longrightarrow 2\Phi(B;0,\sigma^2_{n|n}(f)),\,\,\,\text{where }f(x) = x
\end{align*}
In order to implement the KLPF in practice, one would need bounds on $\Vert x_n - \hat{x}_{n|n}\Vert^2$ and on\\ $P\left(|\hat{x}^{klpf,M}_{n|n} - \hat{x}_{n|n}| > B\right)$ for finite $M$.  

\section{The Separation Principle}
\label{sec: separation}
\begin{figure}
\centerline{\epsfig{figure=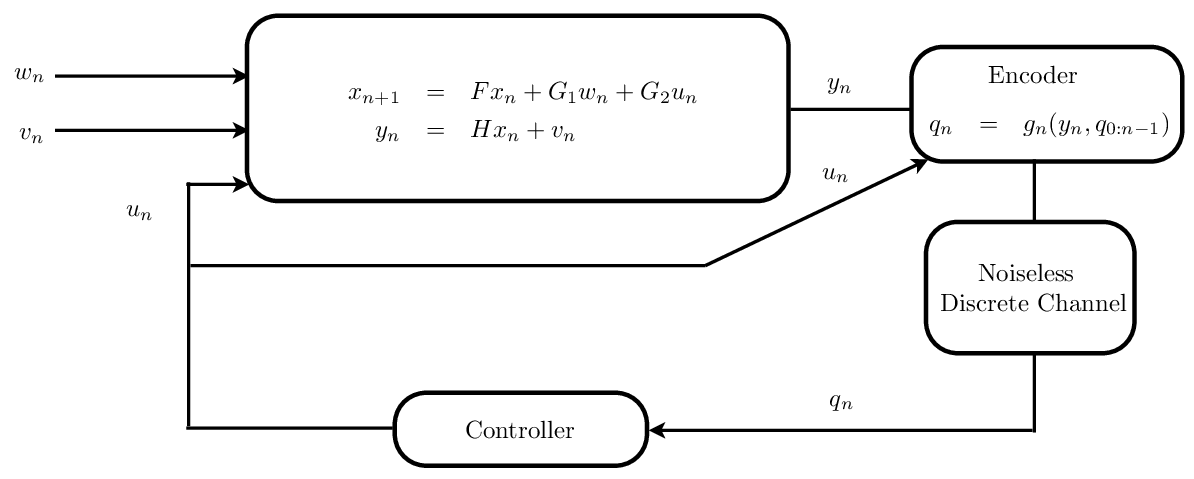, scale=0.65}}
\centering
\caption{Measurement feedback control}
\label{fig: control}
\end{figure}

Consider the closed loop system outlined in Fig \ref{fig: control}.
%
The traditional finite horizon Linear Quadratic Gaussian (LQG) problem \cite{Hinfinity} is one where the control input, $u_{\var}$, is constrained to be a causal and linear function of the measurements $y_{0:n}$, i.e., $u_{\var} = L_n(y_{0},\ldots,y_n)$ for some linear function $L_n(.)$ (or $u_{\var} = L_n(x_{0},\ldots,x_n)$ in the full information case)  and the objective is to minimize a finite horizon quadratic cost function, which can be written as follows
\begin{subequations}
\begin{align}
\min_{\{L_n\}_{0\leq n\leq M}}  &\mathbb{E}_{\{x_{0}, \mathbf{w}_N,\mathbf{v}_N\}} J^c(N),\,\,\,\text{where}\\
J^c(N) &= \sum_{n=0}^N\left[u^T_{\var}M_uu_{\var} + x^T_{\var}M_xx_n\right] + x^T(N+1)M_ox(N+1) \label{eq: cost_function}
\end{align}
\end{subequations}
In the full information case, it is well known that the optimum control action at time $\var$, $u_{\var}$, depends only on the current state $x_{\var}$ and is given by (Ch 9, \cite{Hinfinity})
\begin{subequations}
 \label{eq: full_info}
\begin{align}
u_{\var} &= -K_ux_n,\,\,\, \text{where}\\
K_u &= (M_u + G_2^TM_oG_2)^{-1}G_2^TM_oA\label{eq: feedback_gain}
\end{align}
\end{subequations}
whereas in the case of measurement feedback, the optimal control is given by $u_{\var} = -K_u\hat{x}^{kf}_{\var|\var}$, where $\hat{x}^{kf}_{\var|\var} = Ex_n|y_{0:n}$, which is linear in $y_{0:n}$ due to Gaussianity of the process and measurement noise\footnote{In the absence of Gaussianity, $\hat{x}^{kf}_{n|n}$ would be the linear least mean squared estimate of $x_n$ given $y_{0:n}$}. Note that the control gain in the measurement feedback case is the same as in Eq (\ref{eq: full_info}) and this is the well known separation principle (for eg, \cite{Hinfinity}).

Consider the case when only the quantized measurements $\{q_n\}$ are available and the control action $u_{\var}$ is allowed to be a causal function (not necessarily linear) of the quantized measurements, i.e., $u_{\var} = f_n(q_{0},\ldots,q_n)$, where $f_n(.)$ is any function measurable w.r.t the sigma field generated by $q_{0:n}$. Consider the following control problem
\begin{align}
\label{eq: control_quantized}
\min_{\{f_n\}_{0\leq n\leq M}}  E_{\{x_{0}, \mathbf{w}_N,\mathbf{v}_N\}} J^c(N)
\end{align}
Note that the encoder/quantizer is fixed and the above minimization is over all possible control actions that are causal and measurable functions of the encoder outputs. 
\begin{thm}[The Separation principle]
\label{thm: separation}
 The solution to (\ref{eq: control_quantized}) is given by the following certainty equivalent control law
\begin{align}
u_{\var} = -K_uEx_n|q_{0:n}
\end{align}
where $K_u$, given by (\ref{eq: feedback_gain}), is the same control gain as in the full information case.
\end{thm}
\begin{proof}
The proof for this more general measurement model is a straightforward generalization of the proof presented in \cite{Hinfinity}.
\end{proof}
Let $\hat{x}_{\var|\var} \triangleq Ex_n|q_{0:n}$ and $\tilde{x}_{\var|\var} \triangleq x_n - \hat{x}_{\var|\var}$. Then under the optimal control action, using the orthogonality of $\hat{x}_{\var|\var}$ and $\tilde{x}_{\var|\var}$, and simple algebra, $E J^c_{\var}$ can be decomposed as follows
\begin{align}
EJ^c_{\var} &= \underbrace{tr\left(M_oR_{x_{N+1}}\right) + \sum_{n=0}^Ntr\left((K_u^TM_uK_u + M_x)R_{x_n}\right)}_{J_{LQ}} \nonumber\\
 & + \underbrace{E\tilde{x}^T_{N+1|N+1}M_o\tilde{x}_{N+1|N+1} + E\sum_{n=0}^N\tilde{x}^T_{\var|\var}M_x\tilde{x}_{\var|\var}}_{P^c_{e,N}}
\end{align}
$J_{LQ}$ is the cost under full state information and $P^c_{e,N}$ is the cost that depends on the estimation error covariance. So, the LQG problem of (\ref{eq: control_quantized}) reduces to minimizing $P^c_{e,N}$, completely decoupling estimation and control. Hence the problem of joint optimal estimation and control using quantized measurements reduces to one of finding the optimal causal encoding/quantization rule (see \cite{borkar2} for an interesting treatment of the optimal causal quantization problem). The separation result is not surprising and similar observations in the case of full state information at the encoder were made in \cite{tatikonda2}. The separation principle equipped with the Kalman Like Particle Filter constitutes a computationally feasible framework to solve the optimal LQG problem using quantized measurements. 

\section{Simulations}
\label{sec: sims}
In Alg 1, the particles describe the full probability density of the state conditioned on quantized measurements. While in the KLPF, part of the information about the conditional state density is captured neatly by the Kalman filter. So, the particles describe a truncated Gaussian which has a much smaller covariance than the conditional law of the state given the quantized observations.
We give a few examples in this section to demonstrate the effectiveness of KLPF. Table \ref{tab: table} summarizes the highlights from the first two examples

\begin{table}
\caption{Summary of numerical results}
\label{tab: table}
\begin{center}
\begin{tabular}[c]{|c|c|c|c|c|}
 \hline
 & \multicolumn{2}{c}{Example 1} & \multicolumn{2}{|c|}{Example 2}\\
\cline{2-5}
 & 1 bit & 2 bit & 1 bit & 2 bit\\
\hline
1 bit MLQ-KF & No & - & Yes & -\\ 
\hline
2 bit MLQ-KF & - & No & - & Yes\\
\hline
Alg 1 & 2500 & 10000 & 500 & 1000\\
\hline
KLPF & 500 & 100 & 50 & 10\\
\hline
\end{tabular}
\end{center}
\end{table}

\begin{figure}[ht]
 \begin{minipage}[b]{1.0\linewidth}
  \centering
  \centerline{\epsfig{figure=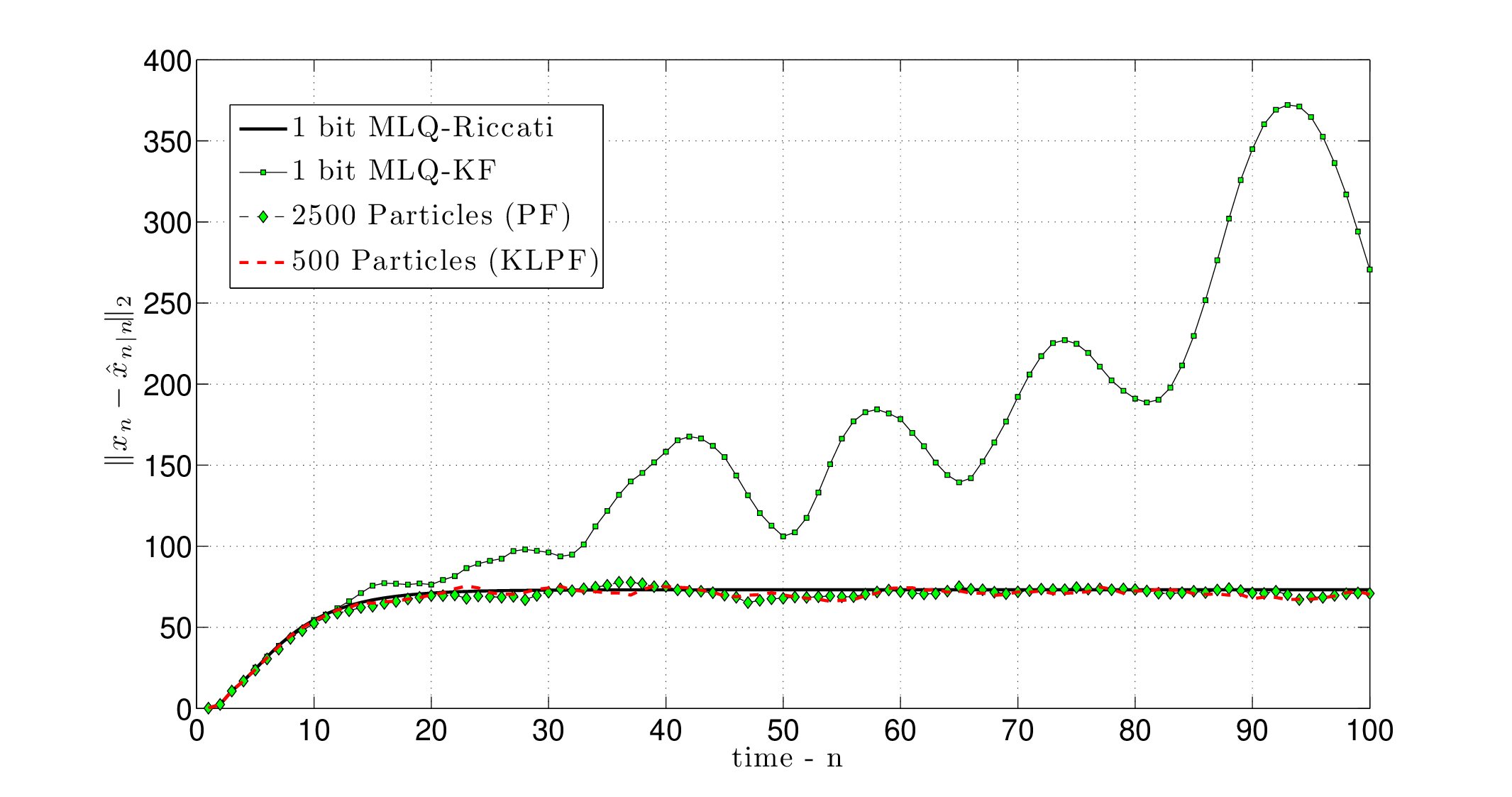, scale=0.3}}
  \centerline{(a) 1 bit MLQ-KF, Alg 1 and KLPF}\medskip
 \end{minipage}
\begin{minipage}[b]{1.0\linewidth}
 \centering
 \centerline{\epsfig{figure=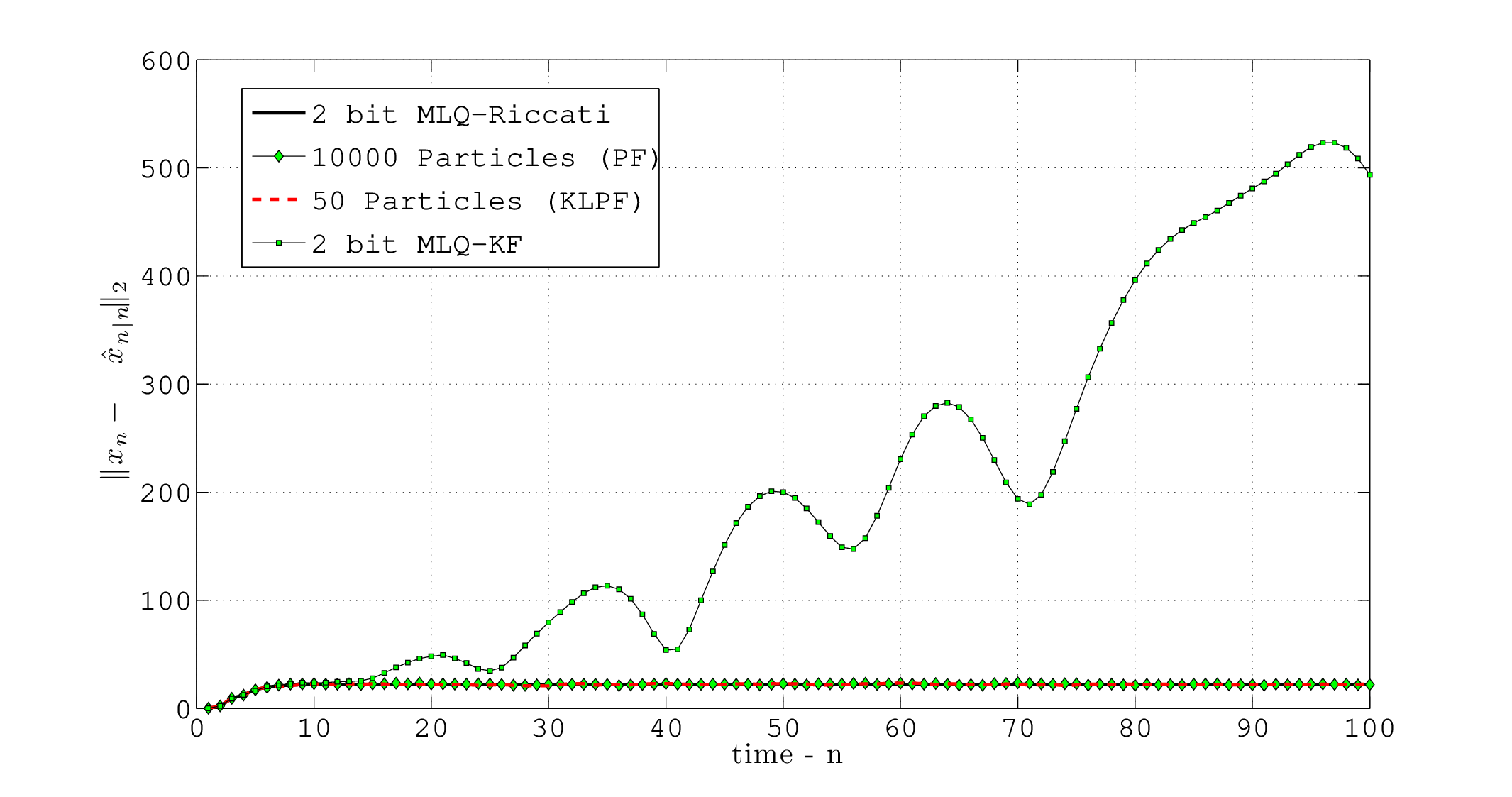, scale = 0.3}}
 \centerline{(b) 2 bit MLQ-KF, Alg 1 and KLPF}\medskip
\end{minipage}
\caption{\small \textit{In (a), 1 bit MLQ-KF clearly diverges while Alg 1 and KLPF converge to the optimal filter. From (b),2 bit MLQ-KF  also diverges while KLPF performs well with just 50 particles}}
\label{fig:example1}
\end{figure}

\begin{figure}[ht]
 \begin{minipage}[b]{1.0\linewidth}
  \centering
  \centerline{\epsfig{figure=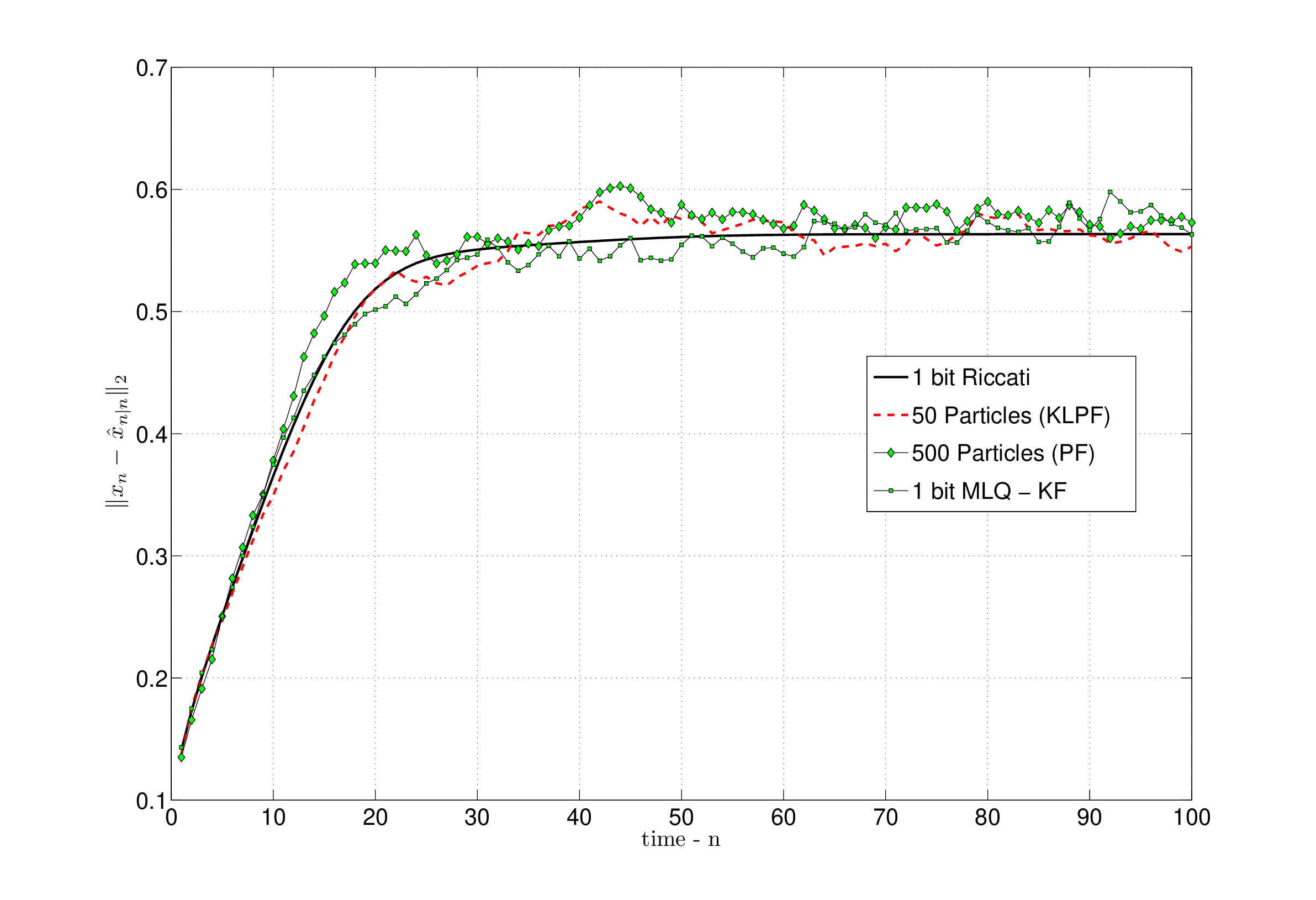, scale=0.25}}
  \centerline{(a) 1 bit MLQ-KF, Alg 1 and KLPF}\medskip
 \end{minipage}
\begin{minipage}[b]{1.0\linewidth}
 \centering
 \centerline{\epsfig{figure=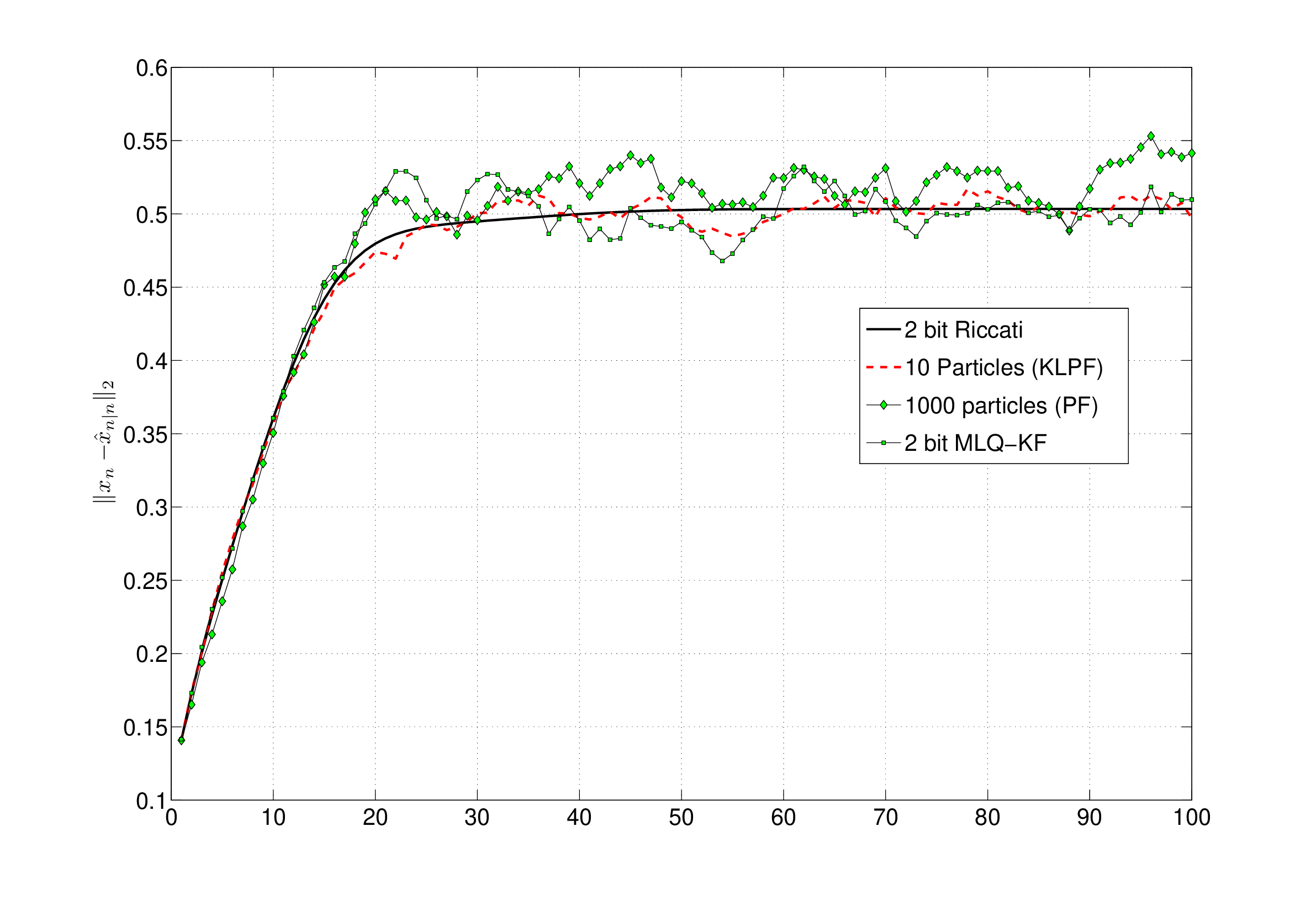, scale = 0.25}}
 \centerline{(b) 2 bit MLQ-KF, Alg 1 and KLPF}\medskip
\end{minipage}
\caption{\small \textit{Both in (a) and (b), all the filters are close to optimal. KLPF achieves good performance with remarkably few particles and hence has a complexity of the same order as that of the MLQ-KF.}}
\label{fig:example2}
\end{figure}
\begin{figure}[ht]
 \begin{minipage}[b]{1.0\linewidth}
  \centering
  \centerline{\epsfig{figure=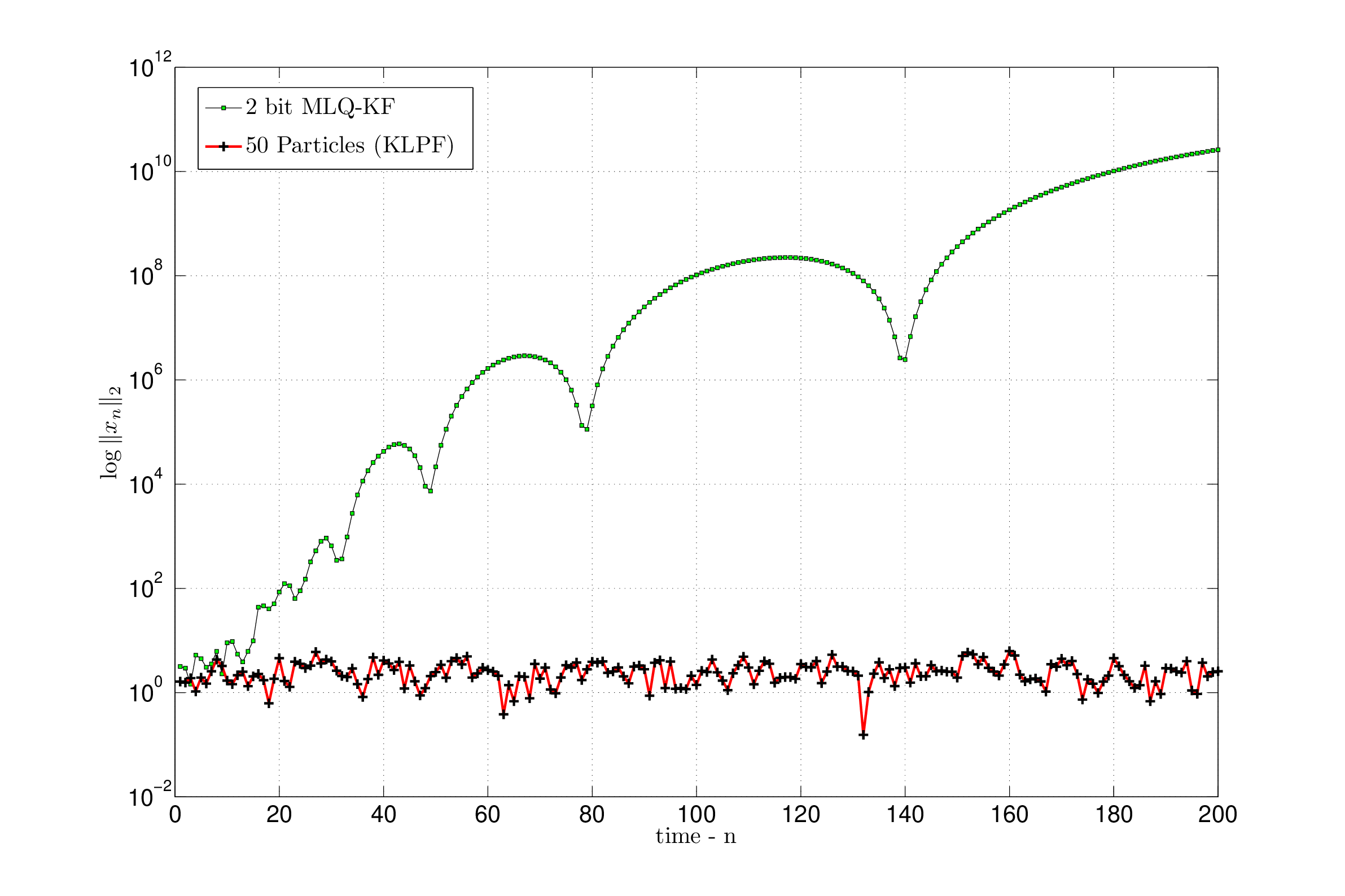, scale=0.25}}
  \centerline{(a) One instance of the system trajectory: x-axis is time and y-axis the magnitude of the state in log scale.}\medskip
 \end{minipage}
\begin{minipage}[b]{1.0\linewidth}
 \centering
 \centerline{\epsfig{figure=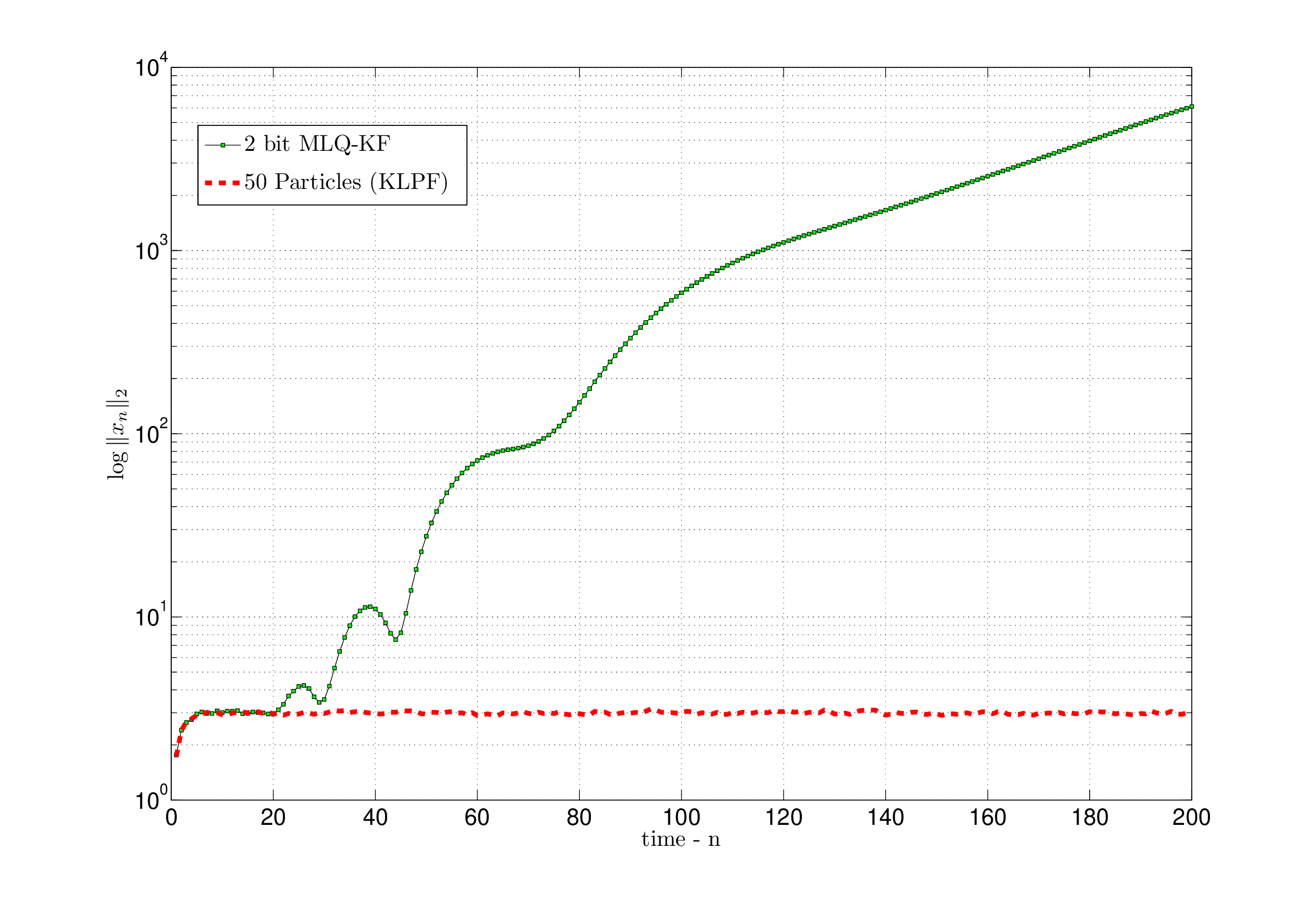, scale = 0.25}}
 \centerline{(b) Averaged over 1000 Monte-carlo iterations.}\medskip
\end{minipage}
\caption{\small \textit{Note that 2 bit MLQ-KF cannot stabilize the system while KLPF can with a hundred particles.}}
\label{fig:example4}
\end{figure}
\begin{figure}[ht]
 \begin{minipage}[b]{1.0\linewidth}
  \centering
  \centerline{\epsfig{figure=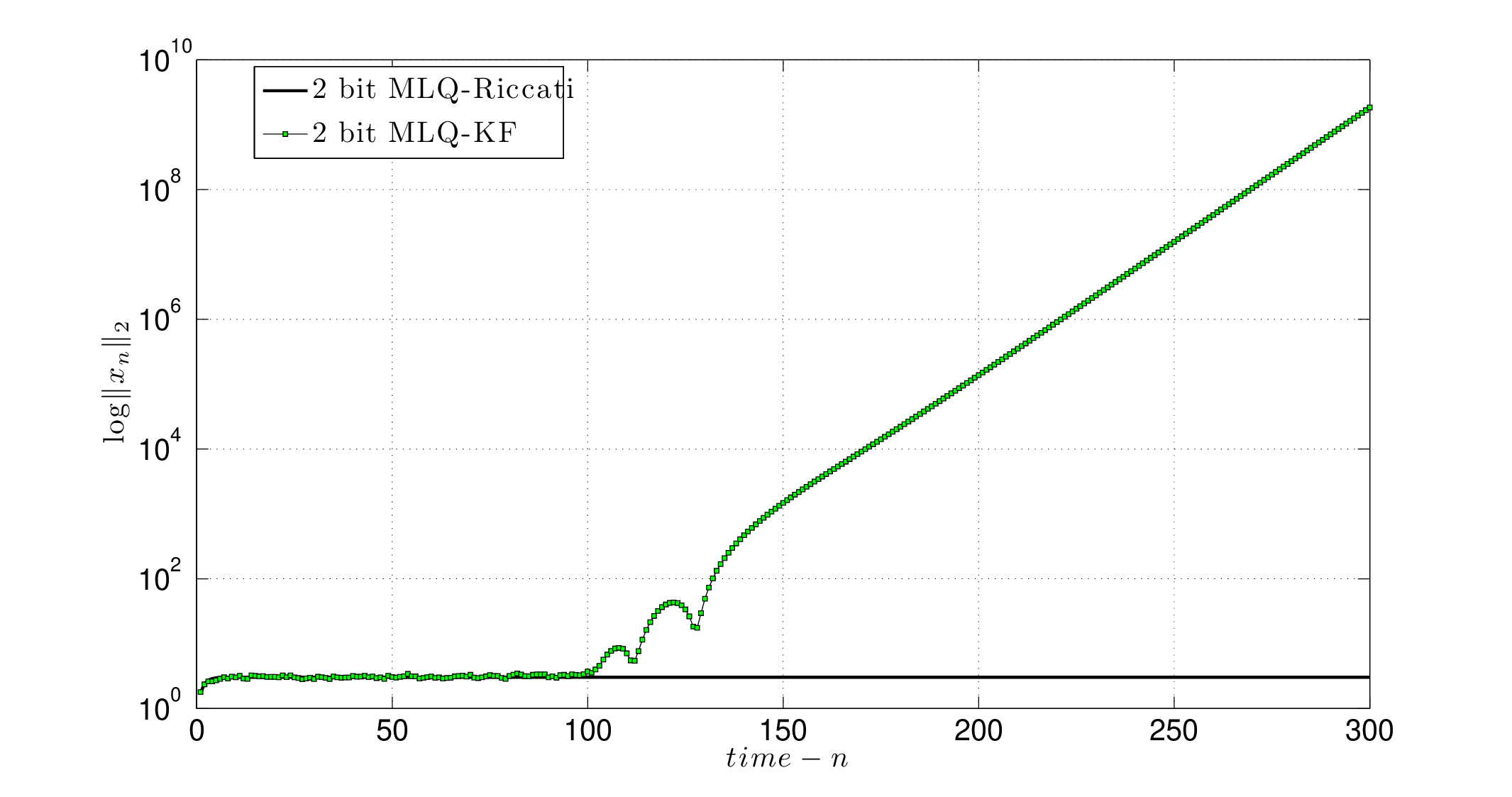, scale=0.25}}
  \centerline{(a) MLQ-KF cannot control the system}\medskip
 \end{minipage}
\begin{minipage}[b]{1.0\linewidth}
 \centering
 \centerline{\epsfig{figure=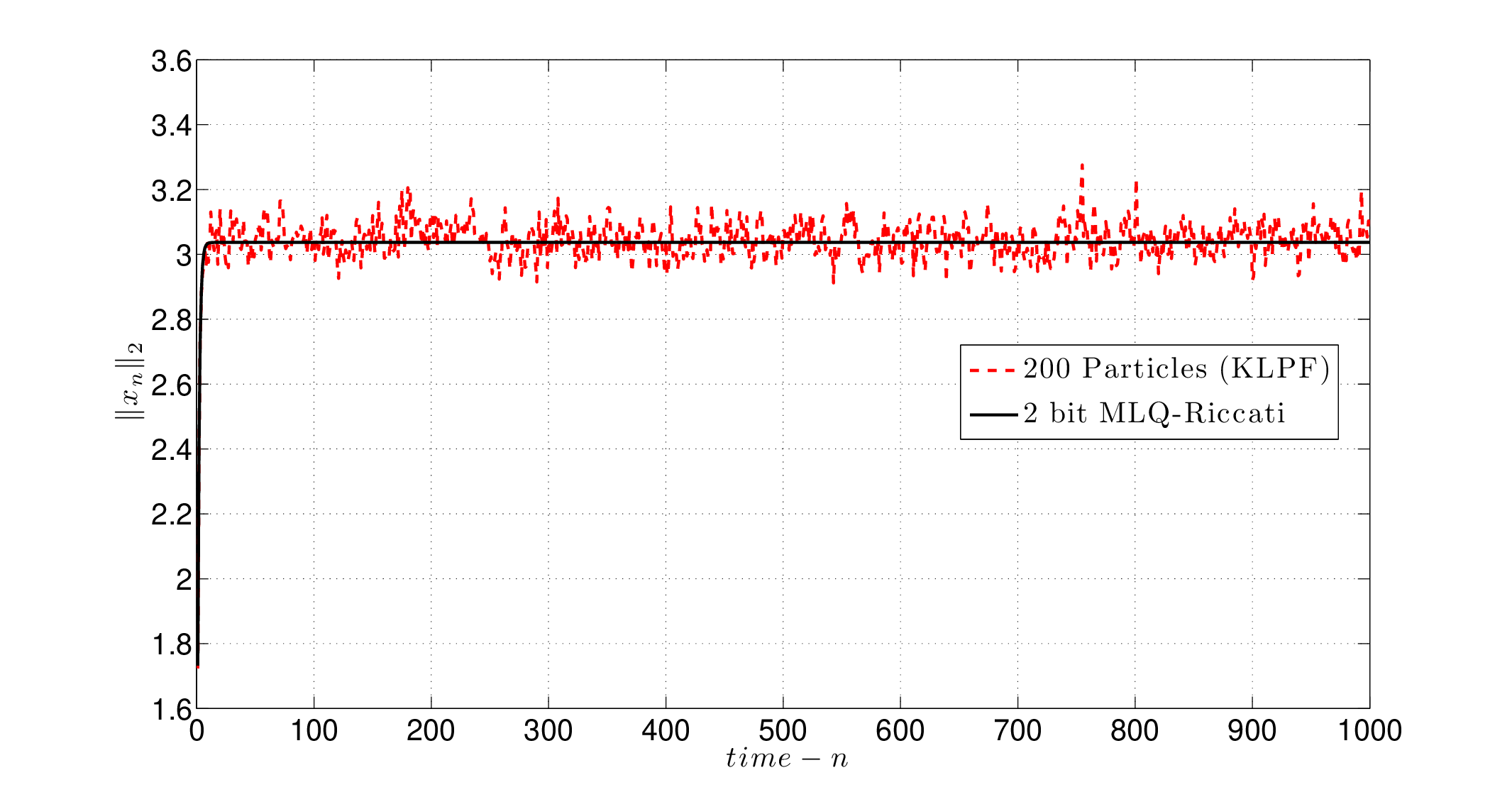, scale = 0.25}}
 \centerline{(b) The KLPF can control the plant.}\medskip
\end{minipage}
\caption{The plot for the KLPF has been shown over a longer time horizon of 1000 time instants to demonstrate convincingly that the KLPF can stabilize the unstable plant.}
\label{fig:example5}
\end{figure}
\begin{figure}[ht] 
\centering
\centerline{\epsfig{figure=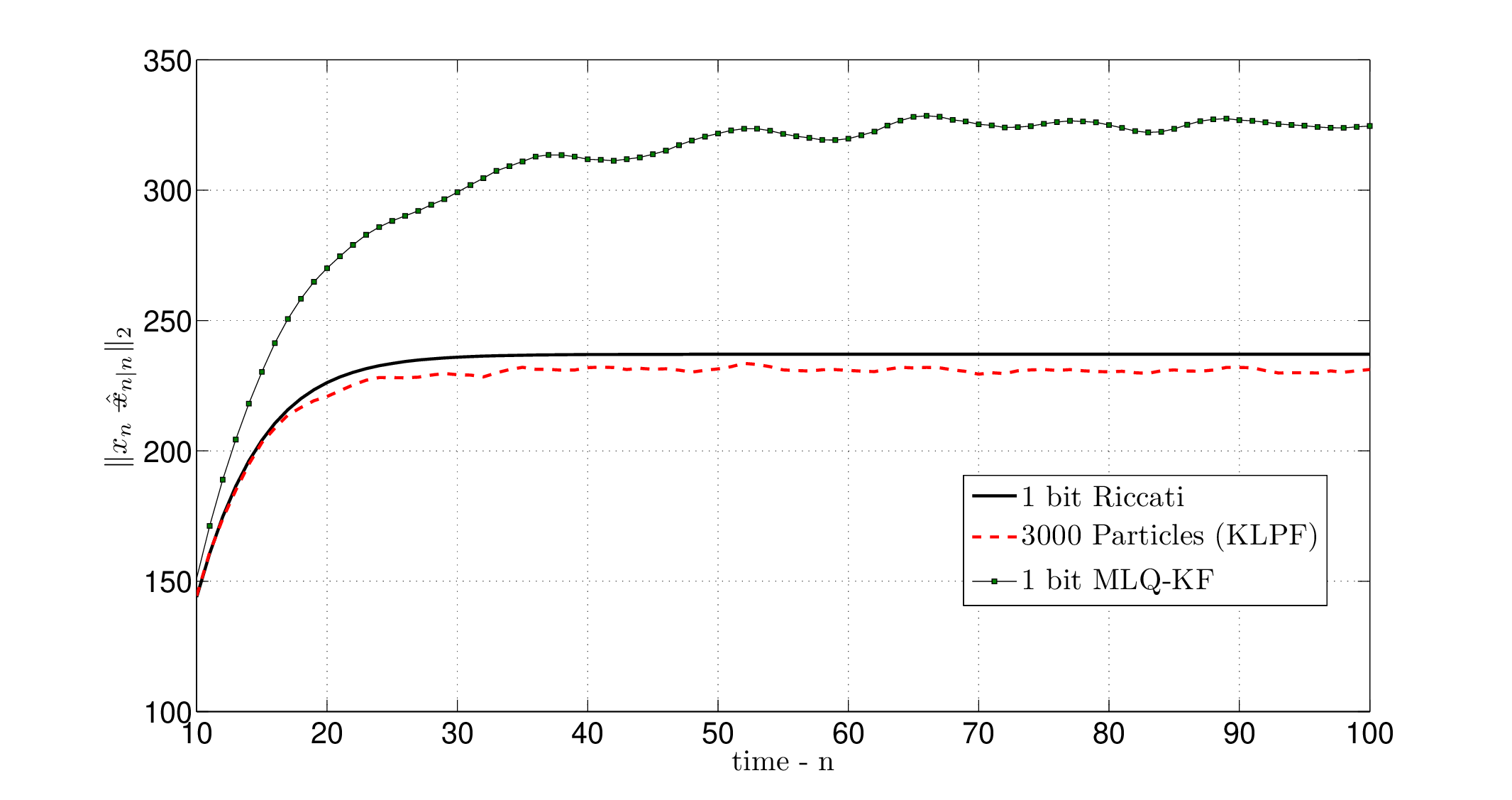, scale=0.25}}
\centerline{(a) Riccati is larger than the optimal error.}
\caption{\small \textit{These plots confirm that the optimal filter does not track the modified Riccati.}}
\label{fig: Example3}
\end{figure}

In Table \ref{tab: table}, a \lq Yes' indicates that the filter works and is close to optimal, a \lq No' indicates that its estimation error diverges and a \lq -' means that the quantization method does not apply to the filter. \lq 1-bit' stands for \lq sign of innovation' and \lq 2-bit' stands for a quantization rule with quantization intervals given by $(-\infty, -1.2437)$, $(-1.2437, -0.3823)$, $(-0.3823, 0.3823)$, $(0.3823, 1.2437)$ and $(1.2437, \infty)$. If the innovation falls in the interval $(-0.3823, 0.3823)$, no measurement update is done, so that 2 bits will suffice to represent the output of the above quantizer. The numbers in front of Alg 1 and KLPF denote the number of particles required to approximate the optimal filter closely. Clearly, KLPF requires far fewer particles than Alg 1. Also evident from Table \ref{tab: table} is the fact that KLPF needs dramatically fewer particles as the quantization becomes finer. 

\textit{Example 1}: Consider a linear time invariant system of the form (\ref{eq: sysdef}) with the following parameters:
$A = \Bigl[\begin{smallmatrix}0.95 & 1 & 0\\0 & 0.9 & 10\\0 & 0 & 0.95\end{smallmatrix}\Bigr]$, $h = [\begin{smallmatrix}1 & 0 & 2\end{smallmatrix}]$, $W = 2\mathcal{I}_3$,  $R = 2.5$ and $P_{0} = 0.01\mathbb{I}_3$, where $\mathbb{I}_m$ denotes an $m\times m$ identity matrix. Note that $A$ is a stable matrix. As can be seen from the plots, 1 bit MLQ-KF and MLQ-KF diverge but KLPF delivers optimal performance with much fewer particles than Alg 1. With the addition of just 1 bit, the required number of particles drops from 500 to 50. 

\textit{Example 2}: A simple tracking system can be characterized by the following parameters, $F = \Bigl[\begin{smallmatrix}1 & \tau\\0 & 1\end{smallmatrix}\Bigr]$, $H = [\begin{smallmatrix}1 & 0\end{smallmatrix}]$, $W = \Bigl[\begin{smallmatrix}\frac{\tau^4}{4} & \frac{\tau^3}{2}\\\frac{\tau^3}{2} & \tau^2\end{smallmatrix}\Bigr]$, $R = 0.81$ and $P_{0} = 0.01\mathbb{I}_3$ and the sampling period $\tau = 0.1$.

In Example 2, note that KLPF works with much fewer particles than in Example 1. One can attribute this to the much higher value of the optimal mean squared error in Example 1 than in Example 2, as can be seen from the plots.

\textit{Example 3}: In \cite{KLPF}, it was noted that the error performance of the optimal filter tracked the modified Riccati and it appeared that the modified Riccati is atleast an upper bound on the error. This was investigated further with more examples and as seen in figure \ref{fig: Example3}, the optimal filter does not track the modified Riccati. This still leaves the possibility that the modified Riccati is an upper bound. Figure \ref{fig: Example3} corresponds to the system defined by $A = \left[\begin{smallmatrix}0.95 &1& 1& 0& 0\\0 &0&.9& 7& 1\\ 0& 0& 0.6& 2& 0 \\0& 0& 0& 0.7& 0.\\ 0& 0& 0& 0& 0.5\end{smallmatrix}\right]$, $h = [\begin{smallmatrix}1 & 0 & 1&0& 2\end{smallmatrix}]$, $W = 2\mathcal{I}_5$,  $R \triangleq R = 2.5$ and $P_{0} = 0.01\mathbb{I}_5$ 

\textit{Example 4 - Closing the loop}: Here, we consider a system for which $x_{n+1} = Fx_n + w_n + u_n$ and $y_n = Hx_n + v_n$, where $F = \Bigl[\begin{smallmatrix} 1&1&1\\0&1&1\\0&0&1\end{smallmatrix}\Bigr]$, $H = [\begin{smallmatrix}1 & 1 & 1\end{smallmatrix}]$, $w_n \sim N_3(0,\mathbb{I}_3)$ is the process noise, $v_n \sim N(0,1)$ is the observation noise and $u_n$ is the control input. Also, consider the quadratic cost function of Eq \eqref{eq: cost_function} with $M_u = 0$, $M_x = M_o = \mathbb{I}_3$. Then the optimal control signal is clearly $u_n = -F\hat{x}_{n|n}$. A seen from fig \ref{fig:example4}, the 2 bit MLQ-KF fails to stabilize the system while KLPF stabilizes it with 100 particles. In fig \ref{fig:example5}, we modify $F$ as $\Bigl[\begin{smallmatrix} 1.1&1&1\\0&1&1\\0&0&1\end{smallmatrix}\Bigr]$ to make it slightly unstable. The plot for the KLPF is over a horizon of 1000 time instants and has been averaged over 1000 monte carlo iterations. The plot indicates that the system trajectory did not diverge in any of the 1000 monte carlo iterations demonstrating that the KLPF can stabilize this unstable system.

\section{Conclusions}
We propose a Kalman like particle filter (KLPF) to optimally track and control a linear Gauss Markov process over a sensor network using quantized measurements. The technique is general and works for an arbitrary causal quantization scheme. In the examples studied, KLPF required moderately small number of particles and therefore can obtain close to optimal performance with a computational complexity comaparable to the conventional Kalman filter. We also showed that the classical separation principle between estimation and control holds. This allowed us to perform optimal LQG control using quantized measurements. 
An important open issue is to determine the number of particles necessary to closely approximate the optimal filter. In order to determine this, one needs upper bounds on the estimation error of the optimal filter and also understand the rate of convergence of particle filters. The error covariance matrix of the optimal filter seems to be upper bounded by the modified Riccati recursion introduced in \cite{bruno}. Determining whether this is the case, and why, remains an interesting open question. In particular, any meaningful upper bound on the estimation error of the optimal filter is necessary for practical applicability of the Kalman Like Particle Filter. 


\small
\bibliographystyle{IEEEbib}
\bibliography{IEEETSP2010}

\appendix
\begin{proof}[Proof of Theorem \ref{thm: state_density}]
\begin{align*}
 p\bra{x_n|q_{0:n}} &= p\bra{x_n}\frac{p\bra{q_{0:n}|x_n}}{p\bra{q_{0:n}}}\\
&= p\bra{x_n}\frac{p\bra{y_{0:n}\in\D^{n+1}|x_n}}{p\bra{q_{0:n}}}\\
&= \phi_d\bra{x_n;0,R_{x_n}}\frac{\Phi_{n+1}\bra{\D^{n+1};R_{y_{0:n},x_n}R_{x_{\var}}^{-1}x_n,\Delta_n}}{\Phi_{n+1}\bra{\D^{n+1};0,R_{y_{0:n}}}}
\end{align*}
Now $R_{y_{0:n},x_n} = \langle y_{0:n},x_n\rangle = \left[\langle\mathcal{Y}_{n-1}, Ax_{n-1}+G_1w_{\var}\rangle,\langle y_n,x_n\rangle\right]^T = \left[ R_{x_n,y_{0:n-1}}, H^T\right]^T$. The recursion for $R_{y_{0:n}}$ follows similarly.
\end{proof}
\begin{proof}[Proof of Theorem \ref{thm: true_innov}]
 Recall the definition of $\xi_{\var}$ from \eqref{eq: kalman_estimate} and note that \eqref{eq: transition} propagates $\xi_{\var}$. Recall that $\{e_n\}$ denotes the innovations process associated to the observation process $\{y_n\}$. So, $e_n = y_n - \E y_n|\mathbf{y}_{n-1} = y_n - HF\xi_{\var-1}$. Now note that $\hat{x}_{\var|\var} \triangleq Ex_n|q_{0:n} = \E\xi_{\var}|q_{0:n}$. So, from \eqref{eq: transition}, we have
\begin{align*}
\hat{x}_{\var|\var} = F\E\xi_{\var-1}|q_{0:n} + K^{f}_n\E e_n|q_{0:n} 
\end{align*}
Since $q_i$ depends only on $e_i$ that is independent of $e_j$ $\forall$ $i\neq j$, we have 
\begin{align*}
\E\xi_{\var-1}|q_{0:n} &= \E\xi_{\var-1}|q_{0:n-1} = \hat{x}_{\var-1|\var-1} \quad \text{and}\\
\E e_n|q_{0:n}  &= \E e_n|q_n = \E e_n|\bigl(e_n\in(z_{l_n},z_{l_n+1})\bigr) \\
&= \Vert e_n\Vert_{_2}\frac{\phi(z_{l_n}) - \phi(z_{l_n+1})}{\Phi(z_{l_n+1}) - \Phi(z_{l_n})} = \sqrt{HP^{kf}_nH^T + R}\frac{\phi(z_{l_n}) - \phi(z_{l_n+1})}{\Phi(z_{l_n+1}) - \Phi(z_{l_n})}
\end{align*}
So, we have 
\begin{align}
\label{eq: Aeq1}
 \hat{x}_{\var|\var} &= F\hat{x}_{\var-1|\var-1} + \displaystyle\frac{P^{kf}_nH^T}{\sqrt{HP^{kf}_nH^T + R}}\frac{\phi(z_{l_n}) - \phi(z_{l_n+1})}{\Phi(z_{l_n+1}) - \Phi(z_{l_n})}
\end{align}
The corresponding error covariance is straightforward using orthogonality. One can rewrite \eqref{eq: Aeq1} as
\begin{align*}
 x_n - \hat{x}_{\var|\var} + \displaystyle\frac{P^{kf}_nH^T}{\sqrt{HP^{kf}_nH^T + R}}\frac{\phi(z_{l_n}) - \phi(z_{l_n+1})}{\Phi(z_{l_n+1}) - \Phi(z_{l_n})} = x_n - \hat{x}_{\var-1|\var-1}
\end{align*}
Using orthogonality of $x_n - \hat{x}_{\var|\var}$ and $\displaystyle\frac{P^{kf}_nH^T}{\sqrt{HP^{kf}_nH^T + R}}\frac{\phi(z_{l_n}) - \phi(z_{l_n+1})}{\Phi(z_{l_n+1}) - \Phi(z_{l_n})}$, the result follows.
\end{proof}
\begin{proof}[Proof of Cor \ref{cor: conv_trueinnov}]
 Under the detectability and stabilizability assumptions, we know that $P^{kf}_n = \Vert x_n - Ex_n|\mathbf{y}_{n-1}\Vert^2$ converges to $P^{kf}$. Let $P^f$ be the steady state value of $P^f_{\var} \triangleq \Vert x_n - Ex_n|y_{0:n}\Vert^2$. Then 
\begin{align*}
 P^{kf} &= FP^fF^T + G_1WG_1^T\\
P^f &= P^{kf} - \frac{P^{kf}H^THP^{kf}}{HP^{kf}H^T + R}
\end{align*}
Also, $\Lambda = F\Lambda F^T + G_1WG_1^T$. Now let $B_n \triangleq P_{\var|\var} - \alpha P^f - (1-\alpha)\Lambda$ and $M_{f,n} \triangleq \displaystyle\frac{P^{kf}_nH^THP^{kf}_n}{HP^{kf}_nH^T + R}$. Also let $M_f$ denote the steady state value of $M_{f,n}$. Then from \eqref{eq: true_innov_recur}, we have
\begin{align*}
 B_n &= FP_{\var-1|\var-1}F^T + G_1WG_1^T - \alpha M_{f,n} - \alpha P^f - (1-\alpha)\Lambda\\
	&= FP_{\var-1|\var-1}F^T + G_1WG_1^T - \alpha M_{f,n} - \alpha \bra{FP^fF^T + G_1WG_1^T - M_f} - (1-\alpha)\bra{F\Lambda F^T + G_1WG_1^T}\\
	&= F\bra{P_{\var-1|\var-1} - \alpha P^f - (1-\alpha)\Lambda}F^T + \alpha\bra{M_f - M_{f,n}}\\
	&= FB_{n-1}F^T + \alpha\bra{M_f - M_{f,n}}
\end{align*}
Since $M_{f,n} \rightarrow M_f$, for each $\epsilon > 0$, there exists an $M$ large enough such that $-\epsilon I \preceq M_f - M_{f,n} \preceq \epsilon I$ for all $n > N$. Then $B_n \rightarrow B$ and $B$ satisfies (Lemma D.1.2 from \cite{Linear_Estimation})
\begin{align}
\label{eq: A2}
 -\epsilon\bra{I + FF^T + F^2(F^T)^2 + \ldots} \preceq B \preceq \epsilon\bra{I + FF^T + F^2(F^T)^2 + \ldots}
\end{align}
Since $F$ is strictly stable and \eqref{eq: A2} is true for each $\epsilon > 0$, $B = \mathbf{0}$. If $F$ is unstable, it is easy to see that $P_{\var|\var}$ diverges to infinity.
\end{proof}

\end{document}